\documentstyle[preprint,aps,eqsecnum,epsfig]{revtex} 
\newcommand{\be}{\begin{equation}}
\newcommand{\ee}{\end{equation}}
\newcommand{\bea}{\begin{eqnarray}}
\newcommand{\eea}{\end{eqnarray}}

\newcommand{\dbarq}{\frac{d^3q}{(2\pi)^3}}
\newcommand{\tilro}{{\tilde{\rho}}}

\newcommand{\bfk}{{\vec k}}
\tighten 
\begin{document} 
%\preprint{PITT-Y2K-1221; LPTHE-00/54} 
%\draft  
\title{\bf NON-FERMI LIQUID ASPECTS OF \\COLD AND DENSE QED AND QCD:
EQUILIBRIUM AND NON-EQUILIBRIUM}   
\author{\bf D. Boyanovsky$^{(a,b)}$, H. J. de Vega$^{(b,a)}$}
\address{(a) Department of Physics and Astronomy, University of 
Pittsburgh, Pittsburgh  PA. 15260, U.S.A\\(b) LPTHE, Universit\'e Pierre et Marie Curie (Paris VI) et Denis Diderot 
(Paris VII), Tour 16, 1er. \'etage, 4, Place Jussieu, 75252 Paris, Cedex 05,France}
\date{\today}
\maketitle 
\begin{abstract} 
We study equilibrium and non-equilibrium aspects of the {\em normal} state of cold and dense QCD and QED.
 The exchange of dynamically screened magnetic gluons (photons) leads to infrared singularities in the 
fermion propagator for excitations near the Fermi surface and the breakdown of the Fermi liquid description. We implement a resummation of these infrared divergences 
via the Euclidean  renormalization group to obtain the spectral density, dispersion relation, widths and
wave function renormalization for single quasiparticles near the Fermi surface. We find that all feature scaling with anomalous dimensions: 
 $\omega_p({k})  \propto  |k-k_F|^{\frac{1}{1-2\lambda}} ~ ; ~ 
\Gamma(k)  \propto  |k-k_F|^{\frac{1}{1-2\lambda}} ~;~  
Z_p({k})  \propto  |k-k_F|^{\frac{2\lambda}{1-2\lambda}}$ with $\lambda  =  
\frac{\alpha}{6\pi} ~ \mbox{for QED}\vspace{0.5 ex} ~,~ 
\frac{\alpha_s}{6\pi} \frac{N^2_c-1}{2N_c} ~~\mbox{for QCD with}~ N_c ~ \mbox{colors and}~ N_F ~ \mbox{flavors}$. The  discontinuity 
of the quasiparticle distribution at the Fermi surface vanishes. For $k\approx k_F$ we find $n_{k \approx k_F} = \frac{\sin[\pi \lambda]}{2\pi \lambda} - \frac{k-k_F}{\pi M(1-4 \lambda)} + {\cal O}(k-k_F)^2 $ with $M$ the dynamical screening scale of magnetic gluons (photons). 
The dynamical renormalization group is implemented to study   non-equilibrium relaxation. The amplitude of single quasiparticle
states with momentum near the Fermi surface falls off as $|\Psi_{k \approx k_F}(t)| \approx |\Psi_{k \approx k_F}(t_0)| e^{-\Gamma(k)(t-t_0)}[t_0/t]^{2\lambda}$. Thus quasiparticle states with Fermi momentum have zero group velocity and relax with a  power law with a coupling dependent anomalous dimension.

\end{abstract} 
%\pacs{12.38.Mh,11.15.-q;11.15.Bt}

\section{Introduction and Motivation} 

There is a substantial theoretical and experimental effort to map the phase diagram of QCD as a function of temperature
(T) and chemical potential ($\mu$). Current theoretical ideas suggest\cite{muzi} that heavy ion collision experiments
from SIS to AGS, SPS, RHIC and finally LHC have the potential to study the region of the phase diagram for $T \leq 300-400~\mbox{Mev}$ and $ \mu \leq  0.6~ \mbox{Gev}$ with higher (T) and lower ($\mu$) for RHIC and LHC. Understanding this
region of the phase diagram can provide insights into the QCD phase transition in the Early Universe, about 10 $\mu s$ 
after the Big Bang as well as the equation of state of hot and dense QCD. Matter at low temperature $ \leq 10~ \mbox{Mev}$ and up to nuclear matter density $\rho_0 \approx 0.16~  \mbox{fm}^{-3}~ ; ~ \mu \approx 300 ~\mbox{Mev}$ is amenable of study by low energy nuclear systems such as multifragmentation phenomena in nuclei. Cold-dense nuclear matter for densities larger than a few times nuclear matter
density cannot be studied with terrestrial accelerators and is the realm of astrophysical compact objects, such as neutron
stars\cite{glendenning,weber}. The fascinanting possibility of detecting a phase transition in quark matter in neutron star X-ray binaries was raised recently\cite{glenweb,blaschke}, where the signal would be a pronounced peak in the frequency distribution of X-ray neutron stars due to a long spin-up stage and the cooling history as revealed by the (soft) X-ray spectra\cite{blas,neutronstars,quarksup}. Thus while QCD at high temperature and relatively small chemical potential can be experimentally studied with ultrarelativistic heavy ion collisions,  astrophysical observation of the properties of neutron stars can provide  observable signatures from cold and dense QCD if quark matter is the
correct  description of the core of spinning neutron stars.  

While there is a substantial body of results on QCD at finite temperature on the  lattice, the lack of manifest reality of the fermion determinant with finite chemical potential presents a problem for
the lattice program with the important exception of two colors which has received considerable attention 
recently\cite{kogut,hands,misha}.

An important aspect of QCD at large chemical potential is that of color superconductivity\cite{bailin} which arises from
a paring instability of the free Fermi gas in the color antitriplet channel. Since the original proposal of color superconductivity via a one (screened) gluon exchange\cite{bailin}, there has been an increased interest in color superconductivity\cite{alford}-\cite{reviews} and diquark condensation\cite{roberts}. The presence of diquark condensates in the cold and dense core of neutron stars could  have potential observable signatures in their cooling history as well as in the magnetic fields of pulsars\cite{blas,neutronstars,quarksup}. Therefore the study of cold-dense QCD is warranted by a definite phenomenological and observable impact if not on terrestrial accelerator experiments certainly in the astrophysical signatures of neutron stars\cite{rajareviu}.

{\bf Goals:} The common framework to study degenerate correlated fermion systems is that of Fermi liquid theory (see next section). The emergence of superconductivity (diquark condensation in the case of QCD) is associated with the instability of the normal Fermi liquid towards an attractive  pairing interaction. In the case of a weakly interacting Fermi system the starting point is the free Fermi gas and pairing results in the opening of a gap in the single particle spectrum at the Fermi surface. Fermi liquid theory is argued to describe the low energy effective field theory of nuclear matter\cite{song} and is therefore an important tool to study nuclear matter and its impact in astrophysical compact objects.
 
Recently it has been argued, within the context of color superconductivity, that the exchange of dynamically screened
(via Landau damping) magnetic gluons results in strong infrared divergences that lead to the breakdown of the Fermi liquid description of cold-dense QCD in perturbation theory\cite{ren}. A similar situation was found in the case of a non-relativistic electron gas with magnetic interactions\cite{holst,reizer,chakra}, where it was argued that there would
be no observable consequences of the breakdown of Fermi liquid theory for {\em terrestrial densities}. 

While diquark condensation and color superconductivity in its various forms have been studied extensively in the literature,
we are aware  of only one previous study of non-Fermi liquid aspects of cold dense QCD\cite{ren}, which revealed large corrections to the superconducting gap\cite{ren}. Further corrections to the superconducting gap were found from lifetime
effects\cite{manuellifetime} not associated with the breakdown of Fermi liquid theory. 

Studying Fermi liquid aspects of the {\em normal} phase is an important part of the program towards understanding cold-dense QCD. As mentioned above, understanding the properties of the normal phase is perhaps the first step towards a complete assessment of the pairing
instabilities and properties of the superconducting state. Furthermore, if the pairing interaction opens a gap at the Fermi surface of
{\em some} quarks, such as the two or three color superconducting phases (2SC or 3SC)\cite{alford}-\cite{rischke,reviews,rajareviu}, the remaining gapless quarks will be described by the concomitant
Fermi liquid. 

Our goal in this article is to provide a comprehensive study of non-Fermi liquid aspects in the {\em normal} phase of cold-dense QCD postponing to a forthcoming article the study of the implications of the breakdown of Fermi liquid theory on color superconductivity. While the study in\cite{ren} focused on the corrections to the color superconducting gap and issues of gauge invariance, we study both equilibrium and non-equilibrium aspects of the non-Fermi liquid behavior. In particular we focus on i) dispersion relation and damping rates of quasiparticles near the Fermi surface, these reveal anomalous dimensions resulting from the breakdown of Fermi liquid theory and ii) the relaxation of these quasiparticles: again we find anomalous relaxation with origin in the same infrared divergences responsible for the breakdown of Fermi liquid theory.  Our main motivations  for initiating this study  and long term goals are manifold: a) to obtain a further  assessment of non-Fermi liquid corrections to the superconducting gaps, critical temperature and spectrum of excitations in the superconducting phase, b) a study of the potential implications of non-Fermi liquid   corrections to the neutrino emissivity and the cooling rate of neutron stars with degenerate quark-matter cores, c) a more complete and detailed understanding of the properties of dense QCD in a regime which is not yet amenable to lattice simulations, d) a study of transport phenomena in the non-Fermi liquid, which is relevant to cooling and thermodynamics of neutron stars. 

In this article we begin this program by studying in detail the breakdown of the Fermi liquid description and by providing a comprehensive analysis of the spectrum of single quasiparticle excitations in the {\em normal} phase along with their relaxation properties. 

{\bf Strategy:} We study both QED and QCD at zero temperature but large (baryon) density so that a perturbative analysis is
reliable. In this regime the hard-dense-loop (HDL) approximation, which is the finite density equivalent of the hard-thermal-loop program of Braaten and Pisarski for finite temperature\cite{brapis,lebellac,blaizot,vanderheyden,manuelbellac,manuelnew}, is reliable and describes the main aspects of static and dynamical screening of gluons (and photons). The leading order in the HDL approximation is the {\em same} in the abelian (QED) and the non-abelian (QCD) theories and screening of gauge fields is completely determined by the one loop quark polarization at finite (and large) 
density\cite{blaizot,vanderheyden,manuelbellac,manuelnew}. In this approximation, static ``longitudinal'' gluons (instantaneous Coulomb interaction) are screened by a Debye screening mass $m_D \propto g\mu$ while transverse gluons are only {\em dynamically}
screened via Landau damping\cite{brapis,lebellac}. To this order in the HDL approximation the polarization tensor for gluons in QCD is similar to that for photons in QED save for trivial color and flavor factors. The main difference between
QCD and QED in this approximation is that while one gluon exchange leads to an attractive (pairing) interaction in the
antitriplet particle-particle channel and therefore to diquark condensation, there is no such attractive channel in QED\footnote{However we expect the Overhauser effect to be present in QED\cite{zahed}, i.e, particles above and holes below the Fermi surface bound by their mutual  (screened) Coulomb attraction. These  are the counterpart of exciton bound states in condensed matter.}. Thus to this order in the HDL approximation, the Fermi liquid aspects of the normal state of cold-dense QCD are similar to those of QED. Thus we  present our study within the framework of QED which accounting for the
trivial color and flavor factors also describe those of the normal state of QCD to leading order in the HDL approximation. 

As described in detail in section II, a Fermi liquid description has an associated ``order parameter'', this is the
jump discontinuity of the Fermi distribution function (of the interacting) system at the Fermi momentum. This discontinuity is given by the residue (wave function renormalization) of the quasiparticle pole for quasiparticles with the Fermi momentum. The breakdown of the Fermi liquid picture is associated with the vanishing of this order parameter, i.e, the
Fermi distribution function is {\em continuous} at the Fermi momentum. We begin by obtaining the quark propagators to leading order in the HDL approximation, corresponding to one bare gluon
exchange in the quark self-energy and show explicitly that to this order there is a sharp discontinuity at the Fermi surface determined by the wave function renormalization of the quasiparticle pole.

However this picture does not survive screening
corrections to the gluon propagator, whereas longitudinal gluon exchange leads to an infrared finite contribution, the
exchange of magnetic gluons which are only {\em dynamically} screened by Landau damping introduce logarithmic divergences in the quark propagator for quasiparticles near the Fermi surface. 

For excitations with  Fermi momentum we argue that these infrared divergences are similar to those of a critical theory
at the upper critical dimensionality. Thus we provide a resummation of the quark propagator for particles with the Fermi 
momentum using the Euclidean renormalization group which reveals the emergence of anomalous dimensions in the spectral
density. Non-equilibrium aspects are studied by implementing a dynamical renormalization group\cite{DRG} which provides a resummation of the quark propagator {\em directly in real time}. The dynamical renormalization group reveals power law relaxation with anomalous dimension for quasiparticles with Fermi momentum.

{\bf Summary of the results:} The exchange of  dynamically screened magnetic gluons leads to infrared divergences in the
single particle propagator for particle excitations near the Fermi surface. These divergences are akin to those found
in critical phenomena for a critical theory at the upper critical dimensionality. We implement a resummation of the perturbative expansion via the Euclidean renormalization group. We find that the particle component of the quark propagator for excitations near the Fermi surface is a {\em scaling function} of the two variables $\tilde{\omega}=\omega-\mu ~;~ \tilde{k}=k-\mu$ with anomalous exponents that depend on the gauge coupling. For $\tilde{k}\neq 0$ the spectral density has the Breit-Wigner (quasiparticle)
form, with the  quasiparticle dispersion relation, width and residue given respectively  by 

\be
\omega_p(\tilde{k})  \propto  |\tilde{k}|^{\frac{1}{1-2\lambda}} ~~; ~~
\Gamma(\tilde{k})  \propto   \sin[\pi \lambda] |\tilde{k}|^{\frac{1}{1-2\lambda}}  ~~; ~~
Z_p(\tilde{k})  \propto  |\tilde{k}|^{\frac{2\lambda}{1-2\lambda}}
\ee
with the effective coupling given by 
\be  
\lambda  =  \left\{
\begin{array}{l}
\frac{\alpha}{6\pi} ~~ \mbox{for QED}\vspace{0.5 ex}\\
\frac{\alpha_s}{6\pi} \frac{N^2_c-1}{2N_c} ~~\mbox{for QCD with}~ N_c ~ \mbox{colors and}~ N_F ~ \mbox{flavors}. 
\end{array}\right.  
\ee

The residue of the quasiparticle pole vanishes as $k \rightarrow k_F$, leading to the following form of the (quasi) particle
distribution function near the Fermi surface

\be
n_{k \approx k_F} = \frac{\sin[\pi \lambda]}{2\pi \lambda} - \frac{\tilde{k}}{\pi M(1-4 \lambda)} + {\cal O}(\tilde{k})^2 ~~; ~~ M= \left\{
\begin{array}{l}
\frac{g \mu}{2 \pi} ~~ \mbox{for QED}\vspace{0.5 ex}\\
\frac{g \mu}{2\pi} \frac{N_F}{2} ~~\mbox{for QCD with}~ N_F ~ \mbox{flavors}. 
\end{array}\right. 
\ee

\noindent revealing the vanishing of the  discontinuity of the Fermi distribution function at the Fermi surface and therefore the vanishing of the Fermi-liquid order parameter.

Implementing a real-time, i.e, a dynamical  version of the renormalization group to study the non-equilibrium relaxation of single quasiparticles near the Fermi surface. We find that the amplitude of the wave function of single quasiparticle states near the
Fermi surface fall off as

\be
|\psi_{k \approx k_F}(t)| \propto |\psi_{k \approx k_F}(t_0)|~e^{-\Gamma(\tilde{k})(t-t_0)} \left[ \frac{t_0}{t} \right]^{2\lambda} \nonumber
\ee

Thus quasiparticles with the Fermi momentum have vanishing group velocity and relax with an anomalous power law. 

We make a comparison between these features of cold-dense QCD (and QED) and those of strongly correlated quasi-one dimensional Fermi systems that feature non-Fermi liquid behavior and are described as Luttinger liquids\cite{schulz,shof}.

The article is organized as follows: in section II we summarize the  aspects of Fermi liquid that are relevant for our
discussion. In section III we obtain the expression for the quasiparticle distribution function in terms of the quark spectral density and obtain the equation of motion for quarks which will be used to study non-equilibrium aspects. 
In section IV we study the equilibrium aspects of single quasiparticles. We begin by studying the quark propagator to lowest order in the HDL approximation, i.e, with the self energy
given by the exchange of hard (bare) gluons and make contact with the Fermi liquid description to this order. 
Soft ($q < g\mu$) gluons require HDL resummation, and the propagator for  particles near the Fermi surface is computed by
including HDL (screening) corrections to the exchanged gluon. The resulting infrared divergences are recognized to be similar to those of a critical theory at its upper critical dimension and resummed using
the Euclidean renormalization group. The renormalization group improved spectral density features scaling behavior that lead to the single quasiparticle dispersion relation and lifetime that scales with anomalous dimensions. We show that the jump discontinuity of the
single particle distribution function vanishes at the Fermi surface as a consequence of the vanishing of the single quasiparticle residue at the Fermi momentum. Section V explores non-equilibrium aspects: the
relaxation of single quasiparticle excitations near the Fermi surface. Implementing a real-time version of the renormalization group reveals that single quasiparticle states with Fermi momentum relax with a power law with anomalous
dimension. In section VI we summarize the connection between QED and the normal state of QCD to the order studied and 
address the important issue of vertex corrections. 
Our conclusions  are discussed in section VII. In this section we also  discuss the striking resemblance of the
spectral density and relaxation to that obtained in a Luttinger liquids\cite{schulz,shof} and some related conjectures on non-Fermi
liquid aspects of high Tc superconductivity, we elaborate on the potential impact of the results and discuss their  range of validity. An appendix contains many of the technical details.

\section{Highlights of a Fermi Liquid}

The most succesful description of interacting degenerate Fermi systems in the {\em normal} state, i.e, non-superconducting or superfluid is Fermi liquid theory. Landau's original formulation, largely phenomenological, has as basic hypothesis
a one-to-one correspondence between the eigenstates of the interacting and non-interacting system. In this formulation
the {\em quasiparticles} are obtained from the single particle states of the non-interacting system via an adiabatic switching-on of the perturbation\cite{schulz,shof,baym}, hence for this picture to remain valid the interactions should not lead
to phase transitions. The {\em Landau quasiparticle} concept is appropriate for excitations very near the Fermi surface,
since for short range interactions the lifetime of these quasiparticles is $\tau \approx |k-k_F|^{-2}$\cite{schulz,shof,baym}, which after including screening effects also holds in the case of Coulomb interactions\cite{ferrel}. For quasiparticles near the Fermi surface the adiabatic hypothesis is reasonable\cite{schulz,shof} and Landau's
phenomenological theory is applicable to study transport phenomena of low energy excitations\cite{baym}. A more modern and consistent
description of Fermi liquid theory is based on renormalization group ideas\cite{shankar} which describe the low energy
physics near the Fermi surface as fixed points of the renormalization group, and Fermi-liquid interactions are those
associated with marginal operators near this fixed point. This formulation of Fermi liquid theory reveals that the physics
near the Fermi surface is very similar to that of critical phenomena and is completely determined by the gapless excitations
associated with the formation of particle-hole states near the Fermi surface\cite{shankar}. 

Fermi liquid theory is the starting point of a consistent study of the properties of degenerate, interacting Fermi systems,
in particular BCS superconductivity can be understood as a result of the instability of the Fermi liquid towards pairing
attractive interactions\cite{schrieffer}. 

It has been adapted to study dense nuclear matter\cite{song} and  recently the renormalization
group description of Fermi liquids has been  extended to the interactions in field theory at  nuclear matter 
density\cite{hsu}. The conclusion of these studies is that Fermi liquid theory is the effective low energy theory for excitations near the Fermi surface. The field
theoretical approach to Fermi liquids quantifies the main concepts of Fermi liquid theory in terms of the spectral
density for the quasiparticles near the Fermi surface: a Breit-Wigner form with a small width (for $k \approx k_F$) and
a finite residue at the ``quasiparticle pole'', $Z_{k_F}$. The quasiparticle distribution function is distorted from the original
Fermi-Dirac step function (at zero temperature) but a ``jump discontinuity'' remains at the Fermi surface which is
determined by the wave function renormalization constant $Z_{k_F}$, (see\cite{daniel} for an explicit calculation in the electron gas). In a well defined sense, this jump discontinuity 
is associated with an ``order parameter'' for a Fermi liquid\cite{shof,schulz}: a normal Fermi liquid corresponds to
$Z_{k_F} \neq 0$ while for a non-Fermi liquid  $Z_{k_F}= 0$. Transport properties of a degenerate Fermi gas depend on the
renormalization of the (quasi) particles, for example the coefficient of the linear power of temperature in the electronic specific heat is proportional to $Z_{k_F}$. 

There are some notable examples of the breakdown of Fermi liquid theory such as the Kondo model of electrons interacting with magnetic impurities, and ``quasi'' one dimensional
metals which provide a novel type of behavior for correlated electron systems: the Luttinger liquid (see\cite{schulz,shof} and references therein). This novel behavior of a correlated degenerate electron system is characterized by a vanishing
jump at the Fermi surface and power law correlations akin to those found in critical phenomena\cite{schulz}.

Another system in which the  Fermi liquid description breaks down is that of non-relativistic electrons 
 interacting via the exchange of ``magnetic'' (transverse) gauge bosons\cite{holst,reizer,chakra}. Recent conjectures
suggest that this type of non-Fermi liquid behavior\cite{varma} or ``Luttinger-liquid'' behavior could explain the unusual
properties of the normal phase of high temperature superconductors\cite{anderson}.

Our study of the normal state of cold-dense QCD (and of QED) reveals a striking resemblance with Luttinger liquid behavior, in particular  the renormalization group improved quark propagator that we find is remarkably similar to that proposed in\cite{varma,anderson} to describe the normal phase of high temperature superconductors.

\section{Preliminaries: quasi particle distribution function and equations of motion}

In this section we obtain the general expressions for the quasiparticle and quasi anti-particle distribution functions
to make contact with Fermi liquid theory. The main aspect of the distribution function is that the existence of a jump discontinuity of the quasiparticle distribution at the Fermi momentum is the signal of Fermi liquid behavior. Furthermore 
we obtain the effective equation of motion for the Dirac field in the medium to extract non-equilibrium aspects. Typically
relaxation es studied by extracting the damping {\em rate} which is obtained from the imaginary part of the self-energy 
evaluated on the mass shell of the Fermion and describes exponential relaxation. However, from the results
 of\cite{blaizot,vanderheyden,manuelbellac,manuelnew} it is
found that the damping rate vanishes for quasiparticles with the Fermi momentum. However the form of the Fermion self-energy 
(see next section) strongly suggests the build-up of logarithmic infrared singularities just like in critical phenomena with the potential for summing up to a power law relaxation with anomalous dimensions. The solution of the real-time equation of motion resummed via the dynamical renormalization group will confirm power law relaxation with anomalous dimension for single quasiparticles with the Fermi momentum. Thus in order to extract the correct relaxation behavior we must study the real-time evolution of the amplitude of the quasiparticle state which is obtained from the  equation of motion.

\subsection{Quasi particle distribution function}

The spatial Fourier transform of the Dirac field operator at any given time $t$ can be written in the form
\begin{equation}
\psi({\vec k},t) = \sum_s \left[b_{{\vec k},s}(t)U^{(s)}({\vec k})+ d^{\dagger}_{{\vec k},s}(t)V^{(s)}(-{\vec k})  \right]
\end{equation} 
with $U^{(s)}({\vec k}); V^{(s)}({\vec k})$ the usual free particle Dirac spinors. We refer to the time dependent operators
$b_{{\vec k},s}(t);d^{\dagger}_{{\vec k},s}(t)$ as the annihilation and creation of quasi-particles and quasi-antiparticles respectively. Within the spirit of Fermi liquid theory, upon adiabatically switching-on of the interaction these operators
interpolate between the free (bare) particles and the dressed (quasi) particles and antiparticles respectively. We define the average number of quasi-particles and quasi-antiparticles
as

\begin{eqnarray}
n_{\vec k} & = & \frac{1}{2}\sum_s \langle b^{\dagger}_{{\vec k},s}b_{{\vec k},s}\rangle \label{particledist}\\
{\bar n}_{\vec k} & = & \frac{1}{2}\sum_s \langle d^{\dagger}_{{\vec k},s}d_{{\vec k},s}\rangle \label{antiparticledist}
\end{eqnarray}

Where the expectation value is in the {\em exact} ground state, which is obtained from the unperturbed ground state
by adiabatically switching on the perturbation from time $t\rightarrow -\infty$ to $t=0$. Using the properties of the usual spinor wave functions $U,V$ and the results of the appendix, it is straightforward to find 

\begin{eqnarray}
n_{\vec k} & = & Tr \left\{\frac{\gamma_0(\not\!{K}+m)\gamma_0}{4\omega_{\vec k}}   
\langle \bar{\psi}({\vec k},t) {\psi}({\vec k},t) \rangle \right\}_{t=0}= Tr \left\{\frac{\gamma_0(\not\!{K}+m)\gamma_0}{4\omega_{\vec k}}\left[-iS^<_{\vec k}(t,t)\right]_{t=0}\right\}
 \label{particledist2}\\
{\bar n}_{\vec k} & = & Tr \left\{\frac{(\not\!{K}-m)}{4\omega_{\vec k}}   
\langle  {\psi}({\vec k},t) \bar{\psi}({\vec k},t) \rangle \right\}_{t=0}=
Tr \left\{\frac{(\not\!{K}-m)}{4\omega_{\vec k}}   
\left[ iS^>_{\vec k}(t,t) \right]_{t=0}  \right\}
 \label{antiparticledist2}
\end{eqnarray}

We are considering the situation of $\mu \gg m$ so that we will neglect the current quark masses and consider the quarks
as massless. Using the spectral representation of the propagators  for Dirac fields given in the appendix, we find for
massless fermions 

\begin{eqnarray}
n_{\vec k} & = & \frac{1}{4} \int dq_0 Tr \left[{\cal P}_+({\vec k})\rho_f(q_0,{\vec k})\right]N_f(q_0) 
 \label{particledistfin}\\
{\bar n}_{\vec k} & = & \frac{1}{4} \int dq_0 Tr \left[{\cal P}_-({\vec k})\rho_f(q_0,{\vec k})\right](1-N_f(q_0)) 
 \label{antiparticledistfin}
\end{eqnarray}

With ${\cal P}_{\pm}({\vec k})$ given in the appendix. Thus to obtain the quasi (anti) particle distribution
functions we need to obtain the fermion spectral density $\rho(q_0,{\vec k})$.

The emission and absorption of hard gluons (photons) with momenta $q \geq \mu$ will affect the distribution functions for fermions from  deep within the fermi sea up to the excitations near the fermi surface. However, because of Pauli blocking of states below the Fermi surface,  the emission and absorption of soft gluons (photons) with $q << \mu$ will only affect the distribution function of particles near the Fermi surface but not those deep within the Fermi sea. Soft gluons (photons) are sensitive to screening corrections and their propagator must include  screening arising from quark loops. The leading order correction is given by the resummation of hard dense loops (HDL) corresponding to quark intermediate states with momenta near the Fermi surface\cite{blaizot,vanderheyden,manuelbellac,manuelnew,lebellac}.  The spectral density in both
cases, with hard and soft gluon exchange, will be studied in detail in the next section.

\subsection{The equation of motion}

We can treat equilibrium and non-equilibrium aspects of cold and dense QED and QCD by studying the real time equation
of motion of a fermion condensate induced by an external source. The equilibrium aspects studied here can all be addressed
by obtaining the fermion propagator and the fermionic spectral density, while the equation of motion allows us to study the
real time relaxation of an initially prepared condensate as an initial value problem prepared by using a suitable source term. The
advantage of studying the equation of motion in real time is that it will reveal the relaxation of fermions directly in
real time. Since the leading order corrections in the HDL approximation
are similar for QED and QCD\cite{blaizot,vanderheyden,manuelbellac,manuelnew,lebellac} and the gluon (photon) polarization is completely determined by one fermion loop, we describe the necessary steps in QED, the final form for the spectral densities and relevant quantities for QCD can be obtained by
simply accounting for the proper color and flavor factors.  

The  QED Lagrangian density in Coulomb gauge is given by (see\cite{DRG} for a similar context)  
 
\begin{eqnarray}
{\cal L}&=& \bar{\Psi}\left(i{\not\!{\partial}}-g \gamma^0 A^0+ g 
\vec{\gamma}\cdot\vec{A}_T\right)\Psi+ \bar{\Psi}\eta+\bar{\eta}\Psi 
\nonumber\\
&&+ \frac{1}{2}\left[\left(\partial^\mu \vec{A}_T\right)^2 
+ \left(\nabla A^0\right)^2\right] \label{lagra}
\end{eqnarray}
where the Grassmann valued source terms were introduced to obtain the
effective Dirac equation in the medium by analyzing the linear response
to these sources.

Following the steps detailed in\cite{DRG}, we find the Dirac equation for the spatial Fourier transform of the expectation value in the massless case to be given by

\begin{equation}
\left(i\gamma_0\frac{\partial}{\partial t}
- {\bbox\gamma}\cdot{\vec k}\right)\psi({\vec k},t) +
\int^{\infty}_{-\infty}dt'\; \Sigma({\vec k},t-t')\;\psi({\vec
k},t')=-\eta({\vec k},t),\label{diraceq1}
\end{equation}
where $\Sigma({\vec k},t-t')$ is the retarded fermion self-energy given by the sum of the transverse and longitudinal
contributions 

\begin{equation}
\Sigma({\vec k},t-t')= \Sigma_T({\vec k},t-t')+\Sigma_L({\vec k},t-t') \label{totsigma}
\end{equation}
Using the results of the appendix we find

\begin{equation}
\Sigma_T({\vec k},t-t') = ig^2 \int \dbarq {\cal P}_{ij}({\bf p}) \left\{
\gamma^i \left[(iS^{++}_{\vec q}(t,t'))(-i{\cal G}^{++}_{T,\vec p}(t,t'))-(iS^{<}_{\vec q}(t,t'))(-i{\cal G}^{<}_{T,\vec p}(t,t'))
\right]\gamma^j
\right\} \label{sigmaT}
\end{equation}

\begin{equation}
\Sigma_L({\vec k},t-t') = ig^2 \int \dbarq  \left\{
\gamma^0 \left[(iS^{++}_{\vec q}(t,t'))(-i{\cal G}^{++}_{L,\vec p}(t,t'))-(iS^{<}_{\vec q}(t,t'))(-i{\cal G}^{<}_{L,\vec p}(t,t'))
\right]\gamma^0
\right\} \label{sigmaL}
\end{equation}

\noindent with ${\vec p}= {\vec k}-{\vec q}$. The propagators are written in terms of their spectral representation and using the results in the appendix it becomes clear that the retarded self-energy has the following causal structure

\begin{equation}
\Sigma({\vec k},t-t')= \Sigma^r({\vec k},t-t')\Theta(t-t')
\end{equation}

Introducing the Fourier transforms in time for the expectation value of the Dirac field, the source and the self-energy,
and  the Fourier  representation of $\Theta(t-t')$  we find that the equation of motion in terms
of the space-time Fourier transforms becomes

\begin{equation} 
\left[\gamma^0 \omega -\vec{\gamma}\cdot {\vec k}+\tilde{\Sigma}(\omega,{\vec k}) \right]\tilde{\psi}(\omega,{\vec k}) = -\tilde{\eta}(\omega,{\vec k}) \label{eqnofmotion}
\end{equation}
Introducing the dispersive representation for the retarded self-energy

\begin{equation}
\tilde{\Sigma}(\omega,{\vec k}) = \frac{1}{\pi} \int dq_0 \frac{\mbox{Im}\tilde{\Sigma}(q_0,{\vec k})}{q_0-\omega-i0^+}\label{dispersion}
\end{equation} 

\noindent  a straightforward calculation reveals that 

\be
\Sigma^r({\vec k},t-t') = \frac{1}{\pi} \int dq_0 e^{-iq_0(t-t')} {\mbox{Im}\tilde{\Sigma}(q_0,{\vec k})} \label{sigmaFT} 
\ee

For massless fermions we write

\begin{eqnarray}
\tilde{\Sigma}(\omega,{\vec k}) & = & \gamma^0 \tilde{\Sigma}_0(\omega,{\vec k})-\vec{\gamma}\cdot\hat{\vec k}\tilde{\Sigma}_1(\omega,{\vec k}) \label{sigmamat}
\end{eqnarray}

Hence the solution of the equation of motion is given by

\begin{equation}
\tilde{\psi}(\omega,{\vec k}) = S_R(\omega,{\vec k}) \tilde{\eta}(\omega,{\vec k}) \label{eqnofmotionsol}
\end{equation}

\noindent with

\begin{eqnarray}
S_R(\omega,{\vec k}) & = &  -\frac{1}{2} \left[{\cal P}_-({\vec k}) S_-(\omega,{\vec k})+ {\cal P}_+({\vec k}) S_+(\omega,{\vec k})  \right] \label{retprop} \\
S_-(\omega,{\vec k}) & = & \left[\omega-k +\tilde{\Sigma}_-(\omega,{\vec k})   \right]^{-1}
\label{Sminus} \\
S_+(\omega,{\vec k}) & = & \left[\omega+k +\tilde{\Sigma}_+(\omega,{\vec k})  \right]^{-1}
\label{Splus} \\
\tilde{\Sigma}_-(\omega,{\vec k}) & = & \tilde{\Sigma}_0(\omega,{\vec k})-\tilde{\Sigma}_1(\omega,{\vec k}) \label{sigmamin}\\
\tilde{\Sigma}_+(\omega,{\vec k}) & = & \tilde{\Sigma}_0(\omega,{\vec k})+\tilde{\Sigma}_1(\omega,{\vec k})\label{sigmaplus}
\end{eqnarray}

\noindent with ${\cal P}_{\pm}({\vec k})$ given in the appendix. The fermion spectral density is obtained from the imaginary
part of the retarded propagator and is given by

\begin{eqnarray}
\rho_f(\omega, {\vec k}) & = &  \frac{1}{2} \left[{\cal P}_-({\vec k})\rho_-(\omega, {\vec k})+ {\cal P}_+({\vec k})\rho_+(\omega, {\vec k})    \right] \nonumber \\
\rho_-(\omega, {\vec k}) & = & \frac{1}{\pi} \frac{\mbox{Im}\tilde{\Sigma}_-(\omega,{\vec k})}{\left[\omega-k +\mbox{Re}\tilde{\Sigma}_-(\omega,{\vec k}) \right]^2+\left[\mbox{Im}\tilde{\Sigma}_-(\omega,{\vec k})\right]^2}\label{rhofmin}\\
\rho_+(\omega, {\vec k}) & = & \frac{1}{\pi} \frac{\mbox{Im}\tilde{\Sigma}_+(\omega,{\vec k})}{\left[\omega+k +\mbox{Re}\tilde{\Sigma}_+(\omega,{\vec k}) \right]^2+\left[\mbox{Im}\tilde{\Sigma}_+(\omega,{\vec k})\right]^2}\label{rhofplus}
\end{eqnarray}

We consider two cases separately to obtain the spectral densities: i) the gluon (photon) line in the
fermion self energy carries {\em hard} spatial momentum $p \geq \mu$. In this case  one (bare) gluon exchange
 gives the leading order correction to the quark self-energy  in the HDL 
approximation\cite{blaizot,vanderheyden,manuelbellac,manuelnew} 
 for large chemical potential.  ii) the gluon (photon)  line carries   {\em soft} spatial
momentum $p \ll \mu$ in which case the gluon (photon) propagator must be dressed by HDL 
fermion loops\cite{blaizot,vanderheyden,manuelbellac,manuelnew}. 
 The contribution from hard gluon exchange to the self-energy of low momentum fermions deep within the Fermi sea, i.e,
with $k << g\mu$,  must be treated non-perturbatively\cite{brapis,blaizot,vanderheyden,manuelbellac,manuelnew} resulting in a modified dispersion relation and their description as quasiparticles. For fermions near the Fermi surface (and weak coupling) the HDL corrections from hard gluon exchange are perturbative. Thus hard gauge fields will modify the fermion propagators for all fermion states in the Fermi sea. 
On the other hand the emission (and absorption) of soft  gluons (photons)  with $q \leq g\mu$ will only affect fermionic
states {\em near the Fermi surface}. 
We now study each case in detail.

\section{Equilibrium aspects}

\subsection{Hard gauge fields (HDL): Fermi-liquid behavior}

We are interested on particles near the Fermi surface, therefore we will focus on the spectral density 
$\rho_-(q_0,\vec{k})$ and we will neglect the anti-particles, for which there is no Fermi surface. 

For hard gauge fields with loop momentum $q \geq \mu$ we can use the free field propagators for transverse and ``longitudinal'' gauge fields given in the appendix. We find that the instantaneous Coulomb interaction leads to a 
local contribution to the self-energy which is subleading for small fermion momentum\cite{manuelbellac,manuelnew}. In the HDL limit we find 

\be
\rho_-(q_0,\bfk)  =   Z_+\delta(q_0 - \omega_+(k))+Z_-\delta(q_0+\omega_-(k)) + \rho_c(q_0,k) \label{spechdl} \\
\ee
\noindent with $\omega_+(k)~,~\omega_-(k)$ the fermion and plasmino quasiparticle poles with residues $Z_{\pm}$
respectively (see\cite{lebellac} for their complete expressions). The 
continuum spectral density has support below the light cone and is given by

\begin{eqnarray}
\rho_c(q_0,k) & = & \frac{1}{\pi} \frac{\mbox{Im}\tilde{\Sigma}_-(\omega,{\vec k})}{\left[\omega-k +\mbox{Re}\tilde{\Sigma}_-(\omega,{\vec k}) \right]^2+\left[\mbox{Im}\tilde{\Sigma}_-(\omega,{\vec k})\right]^2}\label{rhocont}\\
\mbox{Im}\tilde{\Sigma}_-(\omega,{\vec k}) & = & \pi \frac{M^2_f}{2 k}\left(1-\frac{\omega}{k} \right) \Theta(k^2-\omega^2)
\label{imsigmahdl} \\
\mbox{Re}\tilde{\Sigma}_-(\omega,{\vec k}) & = & -\frac{M^2_f}{2 k}\left\{\left(1-\frac{\omega}{k} \right) \ln\left|\frac{\omega+k}{\omega - k } \right| +2  \right\} \label{resigmahdl} \\
M^2_f & = &  \frac{g^2 \mu^2}{8 \pi^2}
\end{eqnarray}
and satisfies the sum rule

\be
Z_+ + Z_- + \int_{-k}^k dq_0 \rho_c(q_0,k) = 1 \label{sumrule}
\ee

The spectral densities $\rho_{\pm}(q_0,\bfk) $ are related by\cite{lebellac} 

\be
\rho_+(q_0,\bfk) = \rho_-(-q_0,\bfk) \label{relation}
\ee

\noindent and upon using  this relation,  the results of the previous section and the appendix we find 

\bea
n_{\bfk} & = & \int dq_0 \rho_-(q_0,\bfk) N(q_0) \label{distpart} \\
\bar{n}_{\bfk} & = & \int dq_0 \rho_-(q_0,\bfk) \bar{N}(q_0) \label{distantipart}\\
{N}(q_0) & = & \Theta(\mu-q_0) ~~; ~~\bar{N}(q_0)  =  \Theta(-\mu-q_0) \label{nbar0}
\eea

\noindent and the fermion number density is given by

\be
{\cal N} = \int \frac{d^3k}{(2\pi)^3} \int^{\mu}_{-\mu} dq_0 \rho_-(q_0,\bfk) 
\ee
\noindent this expression relates the chemical potential to the fermion number density. We now have the tools to understand
the change in the Fermi sea in the HDL limit. 

Since $N(q_0)=\Theta(\mu -q_0)$, and $\omega_+(k) > k$\cite{lebellac}, from the expression for the particle distribution function given by (\ref{distpart}), we see that for values of
$k$ such that $\omega_+(k) < \mu$ the region of support of the spectral density $\rho_-(q_0,k)$ is contained in the
interval $-\infty < q_0 < \mu$, therefore for these values of k the distribution function is $n_k =1$. 
The value of the momentum at which the
frequency of the fermion quasiparticle $\omega_+(k)=\mu$ is the limiting value for which the quasiparticle pole is
in the interval of support of $N(q_0)$, thus defining the Fermi momentum $k_F$ by $\omega_+(k_F)=\mu$.

 We can obtain an estimate for the value of the Fermi momentum in the HDL
limit and weak coupling for which $\mu \gg M_f$  using the large $k$ limit of the quasiparticle pole

\be
\omega_+(k) \approx k + \frac{M^2_f}{k} ~~k \gg M_f \label{largek}
\ee  
\noindent from which we obtain 
\be
k_F \approx \mu \left[ 1-\frac{g^2 }{8 \pi^2}\right]
\ee

 For $k > k_F$ the
quasiparticle pole is outside of the region of support of $N(q_0)$ but the spectral density still has a contribution
inside this region, given by the plasmino pole at $-\omega_-(k)$ and the Landau damping continuum. Hence $n_{k>k_F} \neq 0$ as befits an interacting Fermi system. Thus whereas for $k < k_F$ the distribution function $n_k=1$, for $k>k_F$ it is $0<n_{\bfk} <1$. This analysis leads to the following result 

\be
n_{\bfk}= Z_-(k) + Z_+(k)\Theta(k_F-k)+\int^{\mu}_{-\mu} \rho_c(q_0,k) = \left\{
\begin{array}{l}
1 ~~ k=k_F-\epsilon \vspace{0.5 ex}\\
1-Z_+(k_F) ~~ k=k_F+\epsilon
\end{array}\right. \label{distri}
\ee

We can estimate
the discontinuity or ``jump'' at the Fermi surface in the HDL limit by using the sum rule (\ref{sumrule}) above and the large $k$ limit of the quasiparticle residue\cite{lebellac}   

\be 
1-Z_+(k\approx k_F) \approx \left(\frac{g}{4\pi}\right)^2 \; \left[2 \ln\left(\frac{4\pi}{g} \right) -1
\right] \label{jump}
\ee

At the Fermi momentum we find

\be 
n_{k=k_F} = N(q_0=\mu)Z_+(k_F) + \left(\frac{g}{4\pi}\right)^2 \; \left[2 \ln\left(\frac{4\pi}{g} \right) -1
\right]\label{nkf}
\ee
\noindent where we have used $N(q_0=\mu)=1/2$. 
Thus to this order, the Fermi surface is sharp with a jump discontinuity at $k=k_F$ given by $n_{k=k_F-\epsilon}-n_{k=k_F+\epsilon} = Z_+(k_F)$.  The sharpness of the Fermi surface in the HDL approximation is a consequence that to this order the
quasiparticle excitations have an infinite lifetime.
For momenta $k \approx \mu$ the HDL approximation is not truly justified since in this limit $k \gg g\mu$ and the contribution from the hard momentum region of the two particle cut is of ${\cal O}(g^2 \mu)$ which  becomes comparable to that of Landau damping  of
${\cal O}(g^2 \mu^2 /k)$. This fact notwithstanding, the main purpose of this analysis is to reveal that the contribution
from the hard loop momentum region to lowest order in the resummation program leads to a description consistent with
Fermi liquid theory with a discontinuity or ``jump'' at the Fermi surface determined by the residue of the quasiparticle
pole which to this order is non-vanishing. As we will study in detail below,  the contribution from soft gluon exchange, which requires screening corrections   invalidates the
 Fermi liquid description.

\subsection{Soft gauge fields: Non-Fermi Liquid} 

We now focus on the damping effects on quasiparticles near the Fermi surface, i.e. $k \approx k_F \approx \mu$. Damping
of these excitations results from the emission and absorption of soft gluons which require the gluon (photon) propagators
in the fermion self-energy to be HDL resummed\cite{brapis,lebellac,blaizot,vanderheyden,manuelbellac,manuelnew}. These dressed gauge propagators can be handily included in the calculation of
the self-energy by writing the Wightman functions for the gauge fields in terms of their spectral representation described
in the appendix. Since the fermion momentum is $k \approx k_F \approx \mu$ and the exchanged gluon is soft with $q \ll \mu$, the fermion line in the self-energy does not require HDL resummation and can be taken to be a bare fermion propagator. 

A straightforward calculation using the free fermion propagators for the internal fermion line and the HDL resummed longitudinal and transverse gauge field propagators in terms of their spectral densities leads to the fermion self energy in 
a dispersive representation as in (\ref{dispersion}) with 

\bea
\mbox{Im}\tilde{\Sigma}(q_0,{\vec k})  =  \frac{\pi g^2}{2}\int \dbarq \int dp_0 && \left\{ 
 \left[\Theta(p_0)-\Theta(\mu-q) \right]\delta(q_0-p_0-q)\left[\tilro_T(p_0, p) {\cal P}^{ij}_T({\bf p})\gamma^i {\cal P}_-({\bf q})\gamma^j+  \right. \right.  \nonumber \\
&& \left. \left. \tilro_L(p_0, p) \gamma^0 {\cal P}_-({\bf q})\gamma^0  \right] + \Theta(p_0) \delta(q_0+p_0+q)
 \right. \nonumber \\
 && \left. \left[\tilro_L(p_0, p) {\cal P}^{ij}_T({\bf p})\gamma^i {\cal P}_+({\bf q})\gamma^j+\tilro_L(p_0, p) \gamma^0
{\cal P}_+({\bf q})\gamma^0  \right] \right\} \label{specdensbey}
\eea

\noindent where ${\vec p} = {\vec k - \vec q}$ and we have neglected the instantaneous Coulomb interaction and $\tilde{\rho}_{T,L}$ are given in the appendix.

The dispersive representation (\ref{dispersion}) makes clear that the largest contribution  to the self-energy for
$\omega \approx \mu$ is determined by the behavior of $\mbox{Im}\Sigma(q_0, k)$ for $q_0 \approx \omega \approx \mu$.

The contribution from soft gluons to the self energy of fermions near the Fermi surface $k \approx \mu$ can be extracted easily by first relabelling the integration momenta ${\vec p} \leftrightarrow {\vec k - \vec q}$ so that $\vec q$ is the momentum of the
(soft) gluon line with $q \ll \mu$  and $|{\vec k - \vec q}| \approx k-q\cos(\theta)$  with $\theta$ the angle between $\vec k$ and $\vec q$. In this limit we find

\bea
\mbox{Im}\tilde{\Sigma}_-(q_0, k) & = &  \pi g^2 \int \dbarq \int dp_0 \left\{
\left[ \rho_T(p_0,q)(1-\cos^2(\theta))+\rho_L(p_0,q)\right]\times \right. \nonumber \\
 && \left. \left(\Theta(p_0)-\Theta(\mu-k+q\cos(\theta)) \right)\delta(q_0 - p_0-k+q\cos(\theta))+ \right. \nonumber \\
&& \left. \rho_T(p_0,q)(1+\cos^2(\theta))\Theta(p_0)\delta( q_0+ p_0 +k -q\cos(\theta)) \right\} \label{imsigmaminus} 
\eea

\bea
\mbox{Im}\tilde{\Sigma}_+(q_0, k) & = &  \pi g^2 \int \dbarq \int dp_0 \left\{
\rho_T(p_0,q)(1+\cos^2(\theta))\times \right. \nonumber \\
 && \left. \left(\Theta(p_0)-\Theta(\mu-k+q\cos(\theta)) \right)\delta(q_0 - p_0-k+q\cos(\theta))+ \right. \nonumber \\
&& \left. \left[ \rho_T(p_0,q)(1-\cos^2(\theta))+\rho_L(p_0,q)\right]\Theta(p_0)\delta( q_0+ p_0 +k -q\cos(\theta)) \right\} \label{imsigmaplus} 
\eea

For particles near the Fermi surface the quark propagator has poles near $\omega \approx k \approx \mu$,  while for antiparticles there is no Fermi surface and the poles are at $\omega \approx -k $.  Therefore for particles  the self energy  is determined
by $\mbox{Im}\tilde{\Sigma}_+(q_0, k)$ for $q_0 \approx \omega$ through the dispersion relation. Hence for particles near the Fermi surface the important region is $q_0 \approx \omega \approx k$ which implies that  the argument
of  $\delta(q_0 - p_0-k+q\cos(\theta))$ has support for $p_0 \approx q \cos(\theta)$ corresponding to 
 the region of Landau damping of the gluon (photon) propagator (see appendix). On the other hand for $q_0 \approx k \approx \omega \approx \mu$
the $\delta( q_0 + p_0 +k -q\cos(\theta))$ has
support in the region of $p_0 \approx -2 \mu <0 $  and this contribution is  cancelled by the $\Theta(p_0)$. Thus the
contribution to the self-energy of particles near the Fermi momentum  from soft gluon exchange is completely determined
by the first delta function

  \bea
\mbox{Im}\tilde{\Sigma}_-(q_0, k) & \approx  &  \pi g^2 \int \dbarq \left\{
\left[ \rho_T(p_0,q)(1-\cos^2(\theta))+\rho_L(p_0,q)\right]\times \right. \nonumber \\
 && \left. \left[\Theta(q_0 -k+q\cos(\theta))-\Theta(\mu-k+q\cos(\theta)) \right] \right\}_{p_0  =  q_0-k+q\cos{\theta}} 
\label{imsigmaminusFS} 
\eea

\noindent which obviously vanishes at $q_0=\mu$. Thus we see that the imaginary part of the self energy i.e. the
damping rate for fermionic
excitations vanishes at the Fermi surface\cite{manuelbellac,manuelnew}. For antiparticles the propagator has poles for $\omega \approx -k$, for
hard momentum $k \approx \mu$ and $q_0 \approx -k$ the $\delta(q_0 - p_0-k+q\cos(\theta))$ has support for $p_0 \approx -2k \approx -2\mu$, i.e, in the hard region, while $\delta( q_0+ p_0 +k -q\cos(\theta)) $  has support for $p_0 = -q_0-k+q\cos(\theta) \approx q\cos(\theta)$ which is in the Landau damping region of the spectral density. Hence

  \be
\mbox{Im}\tilde{\Sigma}_+(q_0, k)  \approx    \pi g^2 \int \dbarq  \left\{
\left[ \rho_T(p_0,q)(1-\cos^2(\theta))+\rho_L(p_0,q)\right]\Theta(p_0)  \right\}_{p_0 = -q_0-k+q\cos(\theta)} \label{imsigmaplusFS} 
\ee

The case of antiparticles has been studied in\cite{manuelnew,vanderheyden} with the result that the self-energy is analytic near the Fermi surface and does not lead to novel phenomena. Hence in what follows we will ignore the case of antiparticles and refer the
reader to\cite{manuelnew,vanderheyden} for more details on this case.

 Since the imaginary part of the self-energy for particles vanishes at the Fermi surface, we expand near  $q_0 \approx \mu$ as follows

\be
\Theta(q_0 -k+q\cos(\theta))-\Theta(\mu-k+q\cos(\theta))= \tilde{q_0}\delta(q\cos(\theta)-(k-\mu))+ \cdots
\ee
\noindent with

\be
\tilde{q_0}=(q_0-\mu) \label{tildeq0}
\ee
\noindent and we are led to the following expression

  \be
\mbox{Im}\tilde{\Sigma}_-(q_0, k)  \approx    \frac{ g^2}{4 \pi}\tilde{q_0}  \int_{|k-\mu|}^{q^*}
\left[ \rho_T(\tilde{q_0},q)\left(1-\frac{(k-\mu)^2}{q^2}\right)+\rho_L(\tilde{q_0},q)\right] dq \label{imsigmaminusFS2} 
\ee

\noindent For $\tilde{q_0} \rightarrow 0$ we can approximate

\bea
\rho_T(\tilde{q_0},q) & \approx &  {M^2 ({\tilde{q_0} \over q })\Theta(q^2-\tilde{q_0}^2) \over \left[q^2+4M^2({\tilde{q_0}^2 \over q^2})
 \right]^2+\left[{\pi M^2 \tilde{q_0} \over q }\right]^2 }  \label{rhoTappx} \\
\rho_L(\tilde{q_0},q) & \approx &  {2M^2  ({\tilde{q_0}\over q}) \over \left[q^2+4M^2  \right]^2 } \Theta(q^2-\tilde{q_0}^2) \label{rhoLappx} \\
M^2 & = & \frac{g^2 \mu^2}{4 \pi^2} \label{massglue} 
\eea

The region in which dynamical screening via Landau damping is effective is determined by $q, |\tilde{q_0}| < M$, hence
the validity of the approximation invoked above relies on $|\tilde{q_0}| < M$.

We note that whereas $\rho_T$ has a strong infrared singularity for $q\rightarrow 0$ when $\tilde{q_0} \rightarrow 0$, Debye screening cuts off the infrared behavior of $\rho_L$ which leads to the conclusion that the contribution from the
longitudinal gluons (photons) is $\propto \tilde{q_0}^2$\cite{manuelnew}. 
For excitations at the Fermi surface with $k=\mu$ the contribution to the imaginary part from the transverse
photons can be computed straighforwardly and it yields

\be
\mbox{Im}\tilde{\Sigma}_-^{(a)}(q_0, k) = \frac{g^2}{24\pi}|\tilde{q_0}| \label{imsiga}
\ee 
 
The contribution from the longitudinal gluons (photons) is clearly of ${\cal O}(g^2\tilde{q_0}^2)$ and of the same order as the
contribution from the hard gluon (photon) region\cite{manuelnew}. This is similar to the case of non-relativistic electrons interacting via the screened Coulomb interaction\cite{ferrel}.  The term proportional to $\tilde{k}=k-\mu$ in 
(\ref{imsigmaminusFS2}) can be easily shown to yield a contribution of ${\cal O}(g^2 (\tilde{k}^2/M)
|\tilde{\omega}/M|^{1/3})$.  

Therefore we conclude that

\be
\mbox{Im}\tilde{\Sigma}_-(q_0, k) = \frac{g^2}{24\pi}|\tilde{q_0}| +  A~g^2 \frac{\tilde{q_0}^2}{M}+ B \left(g^2 
\frac{\tilde{k}^2}{M}~\left|{\tilde{\omega}\over M}\right|^{\frac{1}{3}}\right) \label{imsigtot}
\ee
\noindent with calculable coefficients $A,B$. Our focus is to understand the quasiparticle excitations and their
distribution function very near
the Fermi surface, in particular the discontinuity of the distribution function at $k_F$. In order to do this we need
only consider $\tilde{k} \ll M$ with $|\tilde{\omega}| \ll  M$ therefore we will keep only the first (leading) contribution in (\ref{imsigtot}).

We can now obtain the self-energy via the dispersion relation (\ref{dispersion}) for $\omega \approx \mu$ by using the first term of 
(\ref{imsigtot}) and integrating within a region of width $\approx M$ around $\mu$ since this is the region in which
Landau damping is effective for dynamical screening and the region that yields the leading infrared contribution. 

We finally obtain for $k,q_0 \approx \mu$ 

\bea
\tilde{\Sigma}_-(\omega,k=\mu) & = & -\frac{g^2}{24\pi^2} \tilde{\omega} \left\{ \ln\left[-\frac{\tilde{\omega}+i0^+}{M} \right]+
\ln\left[\frac{\tilde{\omega}+i0^+}{M} \right] \right\} \nonumber \\
 & = & -\frac{g^2}{12\pi^2}\tilde{\omega} \ln\left|\frac{\tilde{\omega}}{M}\right|+i \frac{g^2}{24\pi}|\tilde{\omega}| \label{sigmaminfin} \\
\tilde{\omega} & = & \omega-\mu \label{tildew}
\eea

We  finally obtain the retarded  propagator and spectral densities for particles near the Fermi surface  by using the relations (\ref{retprop}-
\ref{rhofplus}) and extracting the term  proportional to ${\cal P}_-({\vec k})$.

It is convenient to introduce $\tilde{k}=  k-\mu \approx k-k_F$ (since from the analysis of the previous section $k_F= \mu(1 - {\cal O}(g^2))$), for $\omega, k \approx \mu$  the inverse propagators for particles can be written as 

\be
S^{-1}_-(\omega,k) \approx  \tilde{\omega} - \tilde{k} +\tilde{\Sigma}_-(\omega,k=\mu)
\ee

\noindent the form of the real part of the self energy $ \propto \tilde{\omega} \ln|\tilde{\omega}/M| $ strongly suggests
 a wave function renormalization. Such an interpretation is obscured by the term $\tilde{k}$ in the propagator, however,
{\em at the Fermi surface} with $\tilde{k}=0$, the logarithmic singularities of infrared origin are reminiscent of those
of a {\em critical theory} and the propagator is similar to that obtained in perturbation theory in a critical theory at the upper critical
dimensionality.
This interpretation is validated by the current ideas on Fermi liquid theory based on the renormalization group, which describes the effective theory near the Fermi surface as a critical theory with   marginal 
Fermi liquid interactions\cite{shankar}. 

Since at the Fermi surface ($\tilde{k}=0$)  the inverse propagator is proportional to $\tilde{\omega}$ the
wave function renormalization constant at the (quasi) particle pole for excitations near the Fermi surface would
be given by

\be
Z(\mu) = \left[ 1+ \frac{d}{d\omega}\mbox{Re}\tilde{\Sigma}_-(\omega,k)|_{\omega=k=\mu} \right]^{-1} \approx
{1 \over {\left[1-\frac{g^2}{12\pi^2}\ln|{\tilde{\omega}\over \mu}|  \right] }}|_{\tilde{\omega}=0}
\ee

This wave function renormalization or residue at the (quasi) particle pole for excitations near the
Fermi surface is precisely the quantity that determines the ``jump'' of the (quasi) particle distribution function
at the Fermi surface as determined by (\ref{nkf}).
 However, the logarithmic singularities manifest in (\ref{sigmaminfin})
would lead to the conclusion that $Z(\mu)=0$ and that the Fermi surface ``vanishes''. Such conclusion has also been obtained
in non-relativistic systems with magnetic interactions\cite{holst,reizer,chakra}. 

A similar situation arises in a critical theory, for example in the Euclidean formulation of a critical scalar theory
with quartic interaction in four dimensions  the inverse propagator for small Euclidean four momentum $K$
is given by 

\be
G^{-1}(K) \approx K^2\left[1-\lambda^2 c \ln(K^2/\kappa^2)\right] \label{scalar}
\ee 
\noindent with $\lambda$ the quartic coupling, $c$ a combinatoric constant and $\kappa$ a renormalization scale. Again the
wave function renormalization or residue at the ``pole'' $K^2=0$, 
\be
Z = \left[\frac{d}{d K^2} G^{-1}(K)|_{K^2=0} \right]^{-1}
\ee

\noindent {\em vanishes}. The vanishing of the wave function renormalization on the particle mass shell in a critical
theory indicates that the propagator acquires an {\em anomalous} scaling dimension. The logarithmic singularities are
resummed via the renormalization group leading to the following improved propagator 

\be
G^{-1}_{RG}(K) = K^2 \left(\frac{K^2}{\kappa^2} \right)^{-\eta} ~~; ~~ \eta= \lambda^2 c
\ee

Therefore we interpret the vanishing of the wave function renormalization for quasiparticles at the Fermi surface as
an indication of the build-up of an anomalous dimension as a consequence of the infrared singularities, which are only
dynamically screened via Landau damping. To make this interpretation explicit we now proceed to obtain a renormalization
group improved quasiparticle propagator, focusing solely on the particle excitations near the Fermi surface.

\subsection{Euclidean Renormalization Group}
\subsubsection{$\tilde{k}=0$} 

In order to obtain a renormalization group resummation of the infrared singularities we must first perform an analitical
continuation to Euclidean space. This is accomplished by the following analytical continuation

\be
\tilde{\omega}+i0^+ = i K \label{euclid}
\ee
\noindent with $K$ taken to be a real variable. From (\ref{Sminus}) the particle propagator now reads  

\bea
S_-(\omega,k)|_{\omega \approx 0,k=0}  & = & -i \frac{K}{\Gamma(K)} \label{sofvert} \\
\Gamma(K) & = &  
{K^2\left[1-\lambda \ln\left[\frac{K^2}{M^2} \right] \right]} ~~ ; ~~ \lambda  =  \frac{g^2}{24\pi^2}
\label{euclidS}
\eea

Obviously $\Gamma(K)$ has the same form as the inverse propagator of a scalar theory with infrared corrections
typical  of a critical theory as discussed above.  The physics near the
Fermi surface requires  $K \ll M$, which obviously leads to a breakdown of the perturbative expansion.
 Just as in a critical theory taking $K \rightarrow 0$ keeping $M$ fixed is the same as keeping
$K$ fixed and taking $M \rightarrow \infty$, i.e, interpreting $M$ as an ultraviolet cutoff and taking the limit of
large cutoff at a fixed transfered momentum.  

The renormalization group improvement proceeds in the same manner as in the scalar theory, we first introduce a wavefunction renormalization constant that will absorb the cutoff dependence  at a given renormalization scale $\kappa$, and
a renormalized vertex function

\bea
\Gamma_R(K,\kappa) & = &  Z[\kappa,M]\Gamma(K,M) \label{renorvertex}\\
Z[\kappa] & = & 1+ \lambda \ln\left[ \frac{\kappa^2}{M^2} \right] + \cdots \label{Zeta}
\eea

The independence of the bare vertex upon the renormalization scale leads to the renormalization group equation

\bea
&& \left[\kappa \frac{\partial}{\partial \kappa} - \eta \right] \Gamma_R(K,\kappa) = 0 \label{RGeq} \\
&&\eta = \frac{\kappa}{Z}\frac{\partial Z}{\partial \kappa} = 2 \lambda \label{eta} 
\eea

Using the fact that the vertex has scaling dimension $2$ i.e,
\be
\Gamma_R(K,\kappa) = \kappa^2 \Phi\left(\frac{K}{\kappa}\right) 
\ee

\noindent with $\Phi$ a dimensionless function of its argument, the renormalization group equation for the scaling function $\Phi$ can be solved straightforwardly leading
to 

\be
\Gamma_R(K,\kappa) = \kappa^2 \left[\frac{K}{\kappa}\right]^{2-\eta} \Phi(1) \label{RGimproved}
\ee

We can now obtain a renormalization group improved vertex function as follows. 
Requiring that at the scale $K=M$ (i.e, at the scale of the cutoff where perturbation theory is valid) the ``bare'' vertex obeys

\be
\Gamma(K=M,M) = M^2 = Z^{-1}[\kappa, M]\Gamma_R[K=M,\kappa]\label{barever}
\ee
\noindent fixes the {\em renormalization group resummed} wave function renormalization in terms of the solution of
the renormalization group equation (\ref{RGimproved}) at the scale $K=M$. The renormalization group improved
``bare vertex'' $\Gamma(K,M) =  Z^{-1}[\kappa, M]\Gamma_R[K,\kappa]$ is valid for $K\ll M$ and  leads to the following resummed particle propagator valid for $k = k_F$,  $\tilde{\omega} < M$ and
weak coupling $\lambda << 1$, 
\be
S_-(\omega,k)|_{\tilde{\omega} \approx 0,\tilde{k}=0} = -\frac{i}{M} \left[\frac{K^2}{M^2} \right]^{\lambda-\frac{1}{2}}\label{SRG}
\ee

The resummed spectral density for particles $\rho_-(\omega,k)= \mbox{Im}S_-(\omega,k)/\pi$ is obtained by performing the
analitic continuation $K\rightarrow  -i\tilde{\omega}+0^+$ leading to the following expression near the Fermi surface

\be
\rho_-(\omega,k)|_{\omega \approx \mu,k=\mu} = \frac{\sin[\pi \lambda]}{\pi |\tilde{\omega}|} \left|\frac{\tilde{\omega}}{M} \right|^{2\lambda} ~~; ~~ \tilde{\omega}= \omega-\mu ~~ ; ~~ \lambda= \frac{g^2}{24\pi^2} \label{specdensfin} 
\ee

It is straightforward to check that this spectral density becomes a delta function in the limit $\lambda \rightarrow 0$ by
integrating it within a small region around the Fermi surface. This spectral density is remarkably similar to those
found in non-fermi liquid systems such as Luttinger liquids and have been experimentally measured in condensed matter
systems via the X-ray edge singularities at the Fermi surface of some metals\cite{mahan}.

Using the renormalization group improved spectral density (\ref{specdensfin}) we can now obtain the value of the
particle distribution function at the Fermi surface by restricting the integral in (\ref{distpart}) to a region of
 width $M$ near
the Fermi surface  

\be
n_{{\vec k}={\vec k}_F} = \int^M_{-M}d\tilde{\omega}~ \Theta(-\tilde{\omega}) {\sin[\pi \lambda]\over \pi \tilde{\omega}} \left| {\tilde{\omega}\over M } \right|^{2\lambda} = N_{{\vec k}={\vec k}_F} {\sin[\pi \lambda]\over \pi \lambda} \label{nkF}
\ee
\noindent with $N_{\vec k}$ the Fermi-Dirac distribution function and $ N_{{\vec k}={\vec k}_F}=1/2$. The distribution function (\ref{nkF}) obviously attains the
free field limit for $\lambda \rightarrow 0$.

\subsubsection{$\tilde{k} \neq 0$} 

The renormalization group method presented above for the case of $\tilde{k}=0$ can be straightforwardly extended to $\tilde{k}\neq 0$ by recognizing that the propagator for $\tilde{k}\neq 0$ is 
\be
\Gamma(K) = K^2\left[1-\lambda \ln\left[ \frac{K^2}{M^2}\right] \right]+iK\tilde{k}. \label{gammaofk}
\ee

\noindent similar to that obtained in a scalar theory in
the large transfered momentum limit but with a ``mass'' term. This similarity suggests that the
term $i\tilde{k}K$ can be treated just as a mass term in the renormalization program of a scalar theory. Therefore, along with the wave function renormalization (\ref{renorvertex}) we also introduce the following multiplicative renormalization

\be
\tilde{k}_R = Z[\kappa,M] \tilde{k} \label{renork}
\ee
\noindent since the are no infrared divergences associated with $\tilde{k}$ then $Z[\kappa,M]$ is chosen in perturbation theory to be
that of the $\tilde{k}=0$ case discussed above and is independent of $\tilde{k}$. 

The renormalization group equation now becomes

\bea
&& \left[\kappa \frac{\partial}{\partial \kappa} - \eta + \gamma \tilde{k}_R \frac{\partial}{\partial \tilde{k}_R}\right] \Gamma_R(K,\tilde{k}_R,\kappa) = 0 \label{RGeqk} \\
&&\eta = \frac{\kappa}{Z}\frac{\partial Z}{\partial \kappa} = 2 \lambda \label{eta2} \\
&& \gamma = \frac{\kappa}{\tilde{k}_R} \frac{\partial \tilde{k}_R}{\partial \kappa} = 2\lambda \label{gammaf}
\eea
\noindent using the fact that the vertex function has dimension $2$ we write

\be
\Gamma_R(K,\tilde{k}_R,\kappa) = \kappa^2 \Phi\left(\frac{K}{\kappa};\frac{\tilde{k}_R}{\kappa}\right)
\ee

Finally introducing the following variables and ansatze

\bea
&& \frac{K}{\kappa} = e^{-t} ~~; ~~ r=\frac{\tilde{k}_R}{\kappa} \label{variables} \\
&& \Phi(e^{-t},r) =  e^{t(\eta-2)} H[t,r] \label{newfunc} 
\eea

\noindent the function $H[t,r]$ obeys the simple renormalization group equation

\bea
&&\left[\frac{\partial}{\partial t} + \bar{\gamma} r \frac{\partial}{\partial r}  \right]H[t,r]=0 \label{finrg}\\
&& \bar{\gamma} = \gamma-1 = 2\lambda-1 \label{gammarg}
\eea

The solution of this equation is simple

\be
H[t,r]= H[\bar{r}(t)] \label{soln}
\ee

\noindent with $\bar{r}(t)$ the solution of the ordinary differential equation

\be
\frac{d \bar{r}}{dt} = - \bar{\gamma}\bar{r} \Longrightarrow \bar{r}(t)= \bar{r}(0)e^{-\bar{\gamma}t} \label{roft}
\ee

Therefore we find that the renormalization group improved vertex function is given by

\be
\Gamma_R(K,\tilde{k},\kappa) = K^2 \left(\frac{K^2}{\kappa^2}\right)^{-\lambda} H\left[ \frac{\tilde{k}{\cal Z}}{K} \left(\frac{K^2}{\kappa^2}\right)^{\lambda} \right] \label{finsolkneq0}
\ee

\noindent where we have explicitly used the multiplicative renormalization of $\tilde{k}$ and introduced the integration constant ${\cal Z}$ that only depends on $\kappa$ and $M$ to be determined later. 

Comparison with the solution (\ref{RGimproved}) for $\tilde{k}=0$ reveals that $H[0]=\Phi(1)$. As 
in the case $\tilde{k}=0$ we  request that the ``bare'' vertex
coincides with the free field expression at the cutoff scale $K=M$, i.e, 

\be
\Gamma[M,\tilde{k},M]=M^2\left[1+i\frac{\tilde{k}}{M}\right] = Z^{-1}[\kappa,M] \Gamma_R[M,\tilde{k},\kappa]
\ee
 \noindent this condition and the identification $H[0]=\Phi(1)$ determine {\em both} $Z[\kappa,M]$ and ${\cal Z}[\kappa,M]$ by comparing the powers of $\tilde{k}$ . Since $H$ is a function of the {\em scaling variable} $\varphi=\tilde{k}{\cal Z}(K/\kappa)^{2\lambda}$ the
functional form of $H[\varphi]$ is uniquely determined for $\tilde{k} \ll M$. We are thus led to the following unique form of the renormalization group resummed vertex function for $\tilde{k} \ll M ~;~ \lambda \ll 1$

\be
\Gamma(K,\tilde{k}) = K^2 \left(\frac{K^2}{M^2}\right)^{-\lambda} \left[ 1 + \frac{i \tilde{k}K}{K^2 \left(\frac{K^2}{M^2}\right)^{-\lambda}}\right] \label{finalgamma} 
\ee

 This discussion reveals another  manifestation of the critical nature of the theory near the Fermi surface:  the renormalization group improved vertex function is of the {\em scaling form}

\be
\Gamma(K,\tilde{k})=\Gamma(K,0)D\left[ \frac{\tilde{k}}{K^{\Delta}} \right]~~;~~ \Delta= 1-2\lambda \label{scalf}
\ee

 Using the relation between the vertex function and the Euclidean particle
propagator given by (\ref{sofvert}) we find from (\ref{finalgamma})

\be\label{prfres}
S_-(\omega,k)_{\omega \approx \mu,k\approx \mu}  =  \frac{1}{\left[iK \left(\frac{K}{M} \right)^{-2\lambda} - \tilde{k} \right]} ~~; ~~ K = -i\tilde{\omega}+0^+ \label{finalpropa}.
\ee

we can now obtain the spectral density near the Fermi surface by following the
steps detailed in the case $\tilde{k}=0$, leading to 

\be
\rho_-(\omega,k)|_{\omega \approx \mu,k\approx \mu} = \frac{\sin[\pi \lambda]}{\pi} |\tilde{\omega}|\left|\frac{\tilde{\omega}}{M}\right|^{-2\lambda}
\frac{1}{
  \left[\left(\tilde{\omega}\left|\frac{\tilde{\omega}}{M}\right|^{-2\lambda}-\tilde{k} \cos[\pi \lambda] \right)^2+\left(\tilde{k} \sin[\pi \lambda] \right)^2   \right]} \label{specdensfink} 
\ee

\noindent which  vanishes identically at $\tilde{\omega}=0$. 
This spectral density is displayed as a function of $\tilde{\omega}$ for several values of $\tilde{k}$ and $ \lambda$ in figures (1) and (2). 

\subsubsection{Quasiparticles}

Figures (1) and (2) clearly reveal that in weak coupling the spectral density features a narrow resonance for $\tilde{k}\neq 0$, the 
position of which is obtained by the vanishing of the first term in the denominator in (\ref{specdensfink}) which determines the
{\em quasiparticle dispersion relation}. It is given by

\be
\tilde{\omega}_p(\tilde{k}) = \mbox{sign}(\tilde{k}) \left[|\tilde{k}| M^{-2\lambda}\cos[\pi \lambda]  \right]^{\frac{1}{1-2\lambda}}
\label{quasidisp}
\ee

We note that the group velocity of the quasiparticles near the Fermi surface

\be
v_g(\tilde{k}) = \left[M^{-2\lambda}\cos[\pi \lambda]  \right]^{\frac{1}{1-2\lambda}}
\frac{|\tilde{k}|^{\frac{2\lambda}{1-2\lambda}}}{(1-2\lambda)} 
\label{groupvel}
\ee

\noindent {\em vanishes} as $k \rightarrow k_F$. We interpret this novel phenomenon in terms of a collective backflow that surrounds
the quasiparticle.

Near the position of the resonance, i.e, for $\tilde{\omega} \approx \tilde{\omega}_p$ the spectral density can be approximated by
a Breit-Wigner form

\bea
&& \rho_-(\omega,k)|_{\tilde{\omega} \approx \tilde{\omega}_p,k\approx \mu} = Z_p[\tilde{k}]\frac{\cos[\pi\lambda]}{\pi} ~ \frac{\Gamma(\tilde{k})}{(\tilde{\omega}-\tilde{\omega}_p(\tilde{k}))^2+\Gamma^2(\tilde{k})} \label{breitwigner} \\
&& Z_p[\tilde{k}] = \frac{\left|\frac{\tilde{\omega}_p(\tilde{k})}{M} \right|^{2\lambda}}{(1-2\lambda)} \label{Zp} \\
&& \Gamma(\tilde{k}) = Z_p[\tilde{k}] |\tilde{k}| \sin[\pi \lambda] \label{width}
\eea

Therefore the residue of the ``quasiparticle pole'' and the ``quasiparticle width'' vanish near the fermi surface as

\bea
&&Z_p[\tilde{k}] \propto |k-k_F|^{\frac{2\lambda}{1-2\lambda}} \label{znear} \\
&& \Gamma(\tilde{k})\propto |k-k_F|^{\frac{1}{1-2\lambda}}\label{widthnear}
\eea

It is straightforward to confirm that the quark propagator (\ref{finalpropa}) has a complex pole with the real and
imaginary parts given by (\ref{quasidisp}) and (\ref{width}) respectively in the narrow width approximation $\Gamma(\tilde{k})/\tilde{\omega}_p(\tilde{k}) \ll 1$. For this purpose write for the complex pole

\be
K_p = -i\tilde{\omega}_p -\Gamma \label{polevalue}
\ee

\noindent with $\Gamma >0 $ corresponding to damping (this is confirmed a posteriori from the solution). In the narrow
width approximation we can replace

\be
(\tilde{\omega}_p -i \Gamma)^{1-2\lambda} \approx (\tilde{\omega}_p)^{1-2\lambda}[1-(1-2\lambda)~i~ \frac{\Gamma}{\tilde{\omega}_p} + \cdots ]
\ee
\noindent requiring the vanishing of the real and imaginary parts of the denominator of (\ref{finalpropa}) we find  the position of the quasiparticle pole and its width given by (\ref{quasidisp}) and (\ref{width}) respectively.

\subsubsection{Requiem to the Fermi Liquid: the jump of the distribution function at $k_F$ vanishes}

We can now study the behavior of the distribution function near the Fermi surface by using the spectral density 
(\ref{specdensfink}) in the expression (\ref{distpart}). Since the domain of validity of the approximations 
invoked to obtain the spectral density is such that $|\tilde{\omega}|~,~|\tilde{k}| < M$ the integral in the 
variable $\tilde{\omega}$ must be cutoff at the scale $M$. Hence we find

\bea
n_{\vec k} & = &  n_{{\vec k}_F} + \Delta n(\tilde{k}) \label{nofkfin} \\
n_{{\vec k}_F} & = & \frac{\sin[\pi \lambda]}{2\pi\lambda} \label{nofkF} \\
\Delta n(\tilde{k}) & = & \int^0_{-M}d\tilde{\omega}\left[\rho_-(\tilde{\omega},\tilde{k})-\rho_-(\tilde{\omega},0) \right]
\label{deltan}
\eea  

 A detailed calculation reveals that

\bea
\Delta n(\tilde{k}) &=& -{\tilde{k} \over \pi M} \sum_{n=0}^{\infty} {
(-\tilde{k}/M)^n \; \sin\left[ (n+2) \pi \lambda \right] \over n+1 - 2
(n+2)\lambda}  \cr \cr
 &=& -{\tilde{k} \over \pi M} \left[ { \sin 2\pi  \lambda \over 1 - 4
\lambda} - {  \tilde{k} \; \sin 3\pi  \lambda \over M(2 - 6 \lambda)} + {
{\tilde k}^2 \; 
\sin 4\pi  \lambda \over M^2(3 - 8 \lambda)} +  {\cal O}({\tilde k}^3)\right]
\eea
 
\noindent leading to the following form of the single quasiparticle distribution near the Fermi momentum

\be
n_{\vec k \approx {\vec k}_F}  =  \frac{\sin[\pi \lambda]}{2\pi\lambda} -\frac{\tilde{k}}{\pi M (1-4\lambda)} + {\cal O}(\tilde{k}^2) \label{distfunckF}
\ee  

This is one of the important results of this work: the exchange of magnetic gluons, which are only dynamically
screened by Landau damping are responsible for the vanishing of the  discontinuity of the distribution
function at the Fermi surface.
This is clearly a consequence of the result that the wave function
renormalization or residue at the quasiparticle pole vanishes as a power law at the Fermi surface. In Fermi liquid
theory, the  discontinuity at the Fermi surface enters in all thermodynamic response functions, for example in
the coefficient of the linear power of temperature in the specific heat\cite{baym,mahan}, thus the vanishing of the
jump discontinuity will likely result in an anomalous specific heat.

\section{Non-Equilibrium aspects: single quasiparticle relaxation}

\subsection{Dynamical Renormalization Group}

The equation of motion (\ref{diraceq1}) can be written in the form

\begin{equation}
\left(i\gamma_0\frac{\partial}{\partial t}
- {\bbox\gamma}\cdot{\vec k}\right)\psi({\vec k},t) +
\int^{t}_{-\infty}dt'\; \Sigma^r({\vec k},t-t')\;\psi({\vec
k},t')=-\eta({\vec k},t),\label{diraceq12}
\end{equation}

\noindent with 

\be
\Sigma^r({\vec k},t-t')   =  \frac{i}{\pi} \int d\omega \mbox{Im}\tilde{\Sigma}(\omega, {\vec k})e^{-i\omega(t-t')} 
 \label{sigmaoft} 
\ee

In order to study the initial value problem as a perturbative expansion, it proves convenient to write this self-energy
in the following form

\bea
\Sigma^r({\vec k},t-t') & = & \frac{\partial}{\partial t'} F(t-t', {\vec k}) \nonumber \\
F(t-t', {\vec k}) & = & \frac{1}{\pi} \int \frac{d\omega}{\omega} \mbox{Im}\tilde{\Sigma}(\omega, {\vec k})e^{-i\omega(t-t')}
\label{Foft} 
\eea

Using the form of the self-energy given by eqn. (\ref{sigmamat}) we see that the function $F$ can be written as

\bea
F(t-t', {\vec k}) & = & \frac{1}{2} \left[{\cal P}_+({\vec k})F_-(t-t', {\vec k}) +{\cal P}_-({\vec k})F_+(t-t', {\vec k})\right] \nonumber \\
F_{\pm}(t-t', {\vec k}) & = & \frac{1}{\pi} \int \frac{d\omega}{\omega} \mbox{Im}\tilde{\Sigma}_{\pm}(\omega, {\vec k}) e^{-i\omega(t-t')} \label{Fplusmin} 
\eea 

We now integrate by parts the non-local term in (\ref{diraceq12}) and use the fact that the adiabatic switching-on of the external
Grassman current leads to a vanishing time derivative  for the fermionic expectation value for
$t<0$\cite{DRG}. Upon switching off the external current
at $t=0$ the effective Dirac equation of motion for the induced expectation value for $t>0$ becomes

\be
\left(i\gamma_0\frac{\partial}{\partial t}
- {\bbox\gamma}\cdot{\vec k}\right)\psi({\vec k},t) + F(0,{\vec k}) \psi({\vec k},t)-
\int^{t}_{0}dt'\; F(t-t', {\vec k})\;\dot{\psi}({\vec
k},t')=0,\label{diraceqgreat0}
\ee

A perturbative expansion of this equation is obtained by writing 

\bea
F(t,{\vec k}) & = & g^2 F^{(2)}(t,{\vec k})+ g^4 F^{(4)}(t,{\vec k})+ \cdots \nonumber \\
\psi({\vec k},t) & = & \psi^{(0)}({\vec k},t)+ g^2 \psi^{(2)}({\vec k},t)+g^4 \psi^{(4)}({\vec k},t)+ \cdots \label{perturba}
\eea
\noindent replacing these expansions in (\ref{diraceqgreat0}) leads to a hierarchy of equations, the first two terms
of which are given by 

\bea
\left(i\gamma_0\frac{\partial}{\partial t}
- {\bbox\gamma}\cdot{\vec k}\right)\psi^{(0)}({\vec k},t) & = &  0 \label{zerothorder} \\
\left(i\gamma_0\frac{\partial}{\partial t}
- {\bbox\gamma}\cdot{\vec k}\right)\psi^{(2)}({\vec k},t) & = & - F^{(2)}(0,{\vec k}) \psi^{(0)}({\vec k},t) + \int^{t}_{0}dt'\;
 F^{(2)}(t-t', {\vec k})\;\dot{\psi}^{(0)}({\vec
k},t') \label{secondorder} 
\eea

The solution of the zeroth order equation (\ref{zerothorder}) is given by

\be
\psi^{(0)}({\vec k},t) = \sum_s\left[B_{{\vec k},s}^{(0)} U^s({\vec k})e^{-ikt} +D_{{\vec k},s}^{*(0)} V^s(-{\vec k})e^{ikt}    \right]
\label{zero}
\ee
while the second order equation (\ref{secondorder}) can be solved in terms of the retarded free field Green's function  

\be
{\cal S}({\vec k},t-t') = -\frac{i}{2} \left[{\cal P}_-({\vec k}) e^{-ik(t-t')}+{\cal P}_+({\vec k}) e^{ik(t-t')}   \right]\Theta(t-t') \label{green0}
\ee

\noindent by proposing the form  

\be
\psi^{(2)}({\vec k},t) = \sum_s\left[B_{{\vec k},s}^{(2)}(t) U^s({\vec k}) e^{-ikt}+D_{{\vec k},s}^{*(2)}(t) V^s(-{\vec k})e^{ikt}    \right]
\label{second}
\ee

\noindent with  the coefficients $B_{{\vec k},s}^{(2)}(t)~,~D_{{\vec k},s}^{*(2)}(t)$ being {\em slowly time dependent}.
We focus on the time evolution of an initial state of particles near the Fermi surface, i.e, we set $D_{{\vec k},s}^{*(0)}=0$ and $k\approx k_F$. A straightforward computation leads to the following result

\bea
B_{{\vec k},s}^{(2)}(t) & = & B_{{\vec k},s}^{(2,a)}(t)+B_{{\vec k},s}^{(2,b)}(t) \nonumber \\
B_{{\vec k},s}^{(2,a)}(t) & = & iB_{{\vec k},s}^{(0)}\left[ t \frac{1}{\pi} \int \frac{d\omega}{\omega}~\mbox{Im}~\tilde{\Sigma}_-(\omega,{\vec k}) \right. \nonumber \\
& + & \left.  \frac{k}{\pi} \int \frac{d\omega}{\omega}~\mbox{Im}~\tilde{\Sigma}_-(\omega,{\vec k})\frac{1}{(\omega -k)} \left[
t- {\sin[(\omega-k)t]\over (\omega-k)} \right]\right] \label{imagpart} \\
B_{{\vec k},s}^{(2,b)}(t) & = & B_{{\vec k},s}^{(0)}\left[ - \frac{k}{\pi} \int \frac{d\omega}{\omega}~\mbox{Im}~\tilde{\Sigma}_-(\omega,{\vec k}) {\left[
1- \cos[(\omega-k)t]\right] \over (\omega-k)^2} \right] \label{realpart} 
\eea

Using the formulae in the appendix of ref.\cite{DRG}, we find that in the limit $t\rightarrow \infty$ the two terms in (\ref{imagpart}) can
be combined to yield

\be
B_{{\vec k},s}^{(2,a)}(t)  =  iB_{{\vec k},s}^{(0)}t~ \mbox{Re}~\tilde{\Sigma}_-(\omega=k,{\vec k}) \label{b2a} 
\ee

 The contribution 
$B_{{\vec k},s}^{(2,b)}(t)$ in the $t\rightarrow \infty$ and $k\approx \mu$  can be understood by writing the integral
in eqn. (\ref{realpart}) as follows

\be
I = - \frac{1}{\pi} \int^M_{-M} d\tilde{\omega}~\mbox{Im}~\tilde{\Sigma}_-(\tilde{\omega},\tilde{k}) {\left[
1- \cos[(\tilde{\omega}-\tilde{k})t]\right] \over (\tilde{\omega}-\tilde{k})^2}  ~~,~~ \tilde{\omega}=\omega-\mu~,~\tilde{k}=k-\mu \label{integral}
\ee  

Using the formulae found in the appendix of ref.\cite{DRG} we find for $\tilde{k} \ll M$ and $Mt \gg 1$ that 

 \bea
B_{{\vec k},s}^{(2,b)}(t) & = & B_{{\vec k},s}^{(0)}\left[  -\Gamma_k t -2 \lambda \ln(\bar{M}t) \right] \label{B2b} \\
\Gamma_k & = & \pi \lambda~|\tilde{k}| =\mbox{Im}~\tilde{\Sigma}_-(\tilde{\omega}=\tilde{k}) \label{gammak} \\
\lambda & = & \frac{g^2}{24\pi^2} ~~; ~~ \bar{M} = Me^{\gamma} 
\eea

\noindent and combining the above results leads to the final lowest order perturbative result
\bea
B_{{\vec k},s}^{(2)}(t) & = & B_{{\vec k},s}^{(0)}\left[-i\delta_k t -\Gamma_k t -2 \lambda \ln(\bar{M}t) \right] \label{B2btot} \\
\delta_k & = & 2\lambda \tilde{k} \ln\left|\frac{\tilde{k}}{M}\right| = -\mbox{Re}~\tilde{\Sigma}_-(\tilde{\omega}=\tilde{k}) \label{deltak} 
\eea

\noindent with $\tilde{\Sigma}_-(\tilde{\omega})$ given by (\ref{sigmaminfin}) and $\gamma$ is the Euler-Mascheroni constant. We note that the damping rate $\Gamma_k$ given by eqn. (\ref{gammak}) coincides with the result obtained in ref.\cite{vanderheyden}, but the logarithmic term is a novel result of our analysis and is the dominant contribution for particles at the Fermi surface. 

Then up to second order we find that the coefficient of the positive energy spinors in the expectation value of the
spinor field is given by

\be
B_{{\vec k},s}(t) = B_{{\vec k},s}^{(0)}\left[1- i\delta_k t - \Gamma_k t -2 \lambda \ln(\bar{M}t)  \right] \label{finalB} 
\ee

The linear and logarithmic secular terms in the solution invalidate the perturbative expansion at very long times. The 
dynamical renormalization group introduces a systematic resummation of these secular terms as described in\cite{DRG}. The implementation of the dynamical renormalization group begins by recognizing that the terms in the square bracket in
(\ref{finalB}) can be interpreted as a renormalization of the coefficient $B_{{\vec k},s}^{(0)}$. Thus following the
procedure detailed in \cite{DRG} we introduce the renormalization constant ${\cal Z}(\tau)$ in the form

\be
B_{{\vec k},s}^{(0)}= B_{{\vec k},s}(\tau){\cal Z}(\tau) ~~\; ; \; ~~ {\cal Z}(\tau)= 1 + \lambda z_1(\tau) +\cdots \label{renodrg}
\ee
\noindent with $\tau$ an arbitrary time scale. The coefficient $z_1(\tau)$ is chosen to cancel the secular terms at the
time scale $t=\tau$. Choosing this scale so that the perturbative expansion is still valid we find that the
improved solution is given by

\be
B_{{\vec k},s}(t) = B_{{\vec k},s}(\tau)\left[1-i\delta_k (t-\tau)-  \Gamma_k (t-\tau) -2 \lambda \ln(\frac{t}{\tau})  \right] \label{finalBren}
\ee
\noindent since the scale $\tau$ is arbitrary the invariance of the solution on the choice of this scale, i.e,

\be
\frac{dB_{{\vec k},s}(t)}{d\tau} =0 \label{rgeqn} 
\ee

\noindent leads to the dynamical renormalization group equation, which to order $\lambda$ is given by

\be
\frac{\partial B_{{\vec k},s}(\tau)}{\partial \tau}+ B_{{\vec k},s}(\tau)\left[i\delta_k+ \Gamma_k+  \frac{2\lambda}{\tau}\right]=0 \label{findrgeqn}
\ee

Finally we  choose the arbitrary scale $\tau$ to coincide with $t$ in the solution of (\ref{findrgeqn})\cite{DRG} leading to
the renormalization group improved time dependent amplitude 

\be
B_{{\vec k},s}(t)=B_{{\vec k},s}(t_0)e^{-i\delta_k(t-t_0)} e^{-\Gamma_k(t-t_0)}\left[\frac{t_0}{t} \right]^{2\lambda} \label{finsoldrg}
\ee

Therefore the time evolution of the wave function of quasiparticle states near the Fermi surface is given by

\be
\psi_{k \approx k_F}(t) = \psi_{k \approx k_F}(t_0)e^{-i(k+\delta_k)(t-t_0)} e^{-\Gamma_k(t-t_0)}\left[\frac{t_0}{t} \right]^{2\lambda}
\ee

We see that the oscillation  frequency $k+\delta_k$ coincides with the quasiparticle dispersion relation (\ref{quasidisp}) to lowest order in $\lambda$  while the
``damping rate'' $\Gamma_k$ coincides with the quasiparticle width (\ref{width}) to lowest order in $\lambda$. The novel aspect is the power law relaxation with anomalous dimension which cannot
be extracted by computing the damping rate in an equilibrium formulation. Thus we are led to one of the important results of this article: quasiparticle excitations with the Fermi momentum relax with a power law with an anomalous dimension.

\subsection{Time evolution from the resummed spectral density}
\subsubsection{ $\tilde{k} =0$} 

We  now use the renormalization group improved form of the propagator for particles near the fermi surface to obtain
the real time evolution from the inverse Fourier transform of (\ref{eqnofmotionsol}). An initial value problem is obtained
by introducing an adiabatically switched-on external Grassman source that vanishes at $t=0$ whose Fourier transform
is given by 

\be
\tilde{\eta}(\omega,k) = \frac{k \gamma^0}{\omega-i0^+} \sum_{s}\left[ B_{{\vec k},s} U^s({\vec k})-D^*_{{\vec k},s}V^s(-{\vec k}) \right] 
\ee 

An initial particle state is prepared by choosing $D_{{\vec k},s}=0$ with ${\cal P}_-({\vec k})\tilde{\eta}(\omega,k)=
2\gamma^0 \tilde{\eta}(\omega,k)$. The Fourier transform is obtained by writing $e^{-i\omega t}= e^{-i\mu t}e^{-i\tilde{\omega}t}$ and performing the analytic continuation (\ref{euclid}). For $k=\mu$ the long time behavior is dominated by
the region $K \approx 0$ for which the analytically continued propagator is given by (\ref{SRG}), we find

\be
\psi({\vec k},t)  \buildrel{kt\gg 1 }\over =\sum_{s}
B_{{\vec k},s} \; U^s({\vec k}) \; e^{-i\mu t} \int^{+ i\infty+0}_{-i\infty+0} 
\frac{dK}{2\pi i M} \; e^{Kt} \; \left({K \over M}\right)^{2\lambda-1} =
\psi({\vec k},0) \; { e^{-i\mu t} \over 2 \pi } { \Gamma(2 \lambda) \;
\sin\left(2 \pi \lambda  \right) \over \left( M t
 \right)^{2\lambda} } \; .  \label{timevol}   
\ee  

The power law can be extracted by a simple scaling of the integration variable $K=z/t$ and  is clearly a consequence of the anomalous scaling behavior of the propagator. Thus the time evolution obtained directly from the Fourier transform of the
(Euclidean) renormalization group improved propagator confirms the result from  the dynamical (real time) renormalization group for excitations with Fermi momentum ($\tilde{k}=0$) and relates the power law fall-off to the anomalous scaling dimension. 

\subsubsection{$\tilde{k} \neq 0$}

The Fourier transform with the improved propagator (\ref{finalpropa}) is more complicated because of the  branch cut and
 the complex pole.  Although obtaining an exact expression for the full integral
is a complicated task, we can confirm the exponential relaxation in the narrow width approximation (weak coupling) from the contribution
from the complex pole (\ref{polevalue}). The  real and imaginary parts of which are given by (\ref{quasidisp}) and (\ref{width}) respectively in the narrow width approximation. Keeping only the contribution of the complex pole to the Fourier transform  for $\tilde{k} \neq 0$ we find

\be
\psi({\vec k},t)  \buildrel{kt\gg 1 }\over \approx 
\psi({\vec k},0) e^{-i\mu t} e^{-i\tilde{\omega}_p(\tilde{k})t} e^{-\Gamma(\tilde{k})t} .  \label{timevoldamp}   
\ee  

 The exponential relaxation confirms the result of the dynamical renormalization group since the narrow width approximation
implies $\lambda <<1$ for which the power law arising from the cut contribution is subleading. 

Thus the study of both cases $\tilde{k} =0$ and $\tilde{k}\neq 0$ via the Fourier transform of the renormalization group
improved propagator confirms the results obtained directly in real time via the dynamical renormalization group. As was
emphasized above, the most important novel feature is the power law relaxation with anomalous dimension for quasiparticles
with Fermi momentum as well as the anomalous scaling dimension for the quasiparticle dispersion relation, width and residue.

\section{From QED to QCD} 

Although we have focused our calculations on the case of QED, it is straightforward to extrapolate the results to
QCD to the same order in the HDL approximation. The reason is that the polarization tensor for gluons (photons) to this
order is given by the quark loop. In the cold non-abelian theory, the gluons give only the vacuum contribution while the
quarks dominate the polarization tensor for large density. This is unlike the case at finite temperature where the
gluon loops give a contribution of the same order as that of the quark loop. Furthermore, again unlike finite temperature, a magnetic
gluon mass is {\em not} expected to arise because the gluons do not have any infrared divergence. The important changes
that are required to extrapolate the results from QED to QCD are the following:

{\bf a):} The gluon  mass scale (\ref{massglue}) that enters in the longitudinal and transverse spectral densities 
(\ref{rhoLappx}) and (\ref{rhoTappx}) respectively now becomes
$$
M^2_{QCD} = \frac{g^2 \mu^2}{4\pi^2} \left(\frac{N_F}{2}\right) 
$$
  with $N_F$ the number of flavors of quarks in the fundamental representation.  

{\bf b):}
The effective coupling in the self-energy of the quark now includes the trace over the color matrices resulting in that for QCD 
$$
\lambda_{QCD} = \frac{g^2}{24\pi^2} \frac{N^2_c-1}{2N_c}
$$

Thus the results obtained for QED to leading order in the HDL approximation can be extrapolated to QCD via the
replacement $M \rightarrow M_{QCD}~;~ \lambda \rightarrow \lambda_{QCD}$. 

\vspace{2mm}

{\bf Vertex corrections:} an important question that must be addressed is that of vertex corrections. In leading
order in HDL, QCD and QED are similar because the quark loop gives the leading contribution to the vector boson
polarization tensor, therefore the only relevant vertex is abelian. Since the vertex is related to the self-energy
by the Ward identity one would expect logarithmic corrections arising from  the vertex. However the results obtained
from a detailed study in\cite{ren} show that this is not the case. The basic point of the argument is the following,
the small momentum limit of the vertex, which is the relevant limit for long wavelength gluons, is related to the
quark self-energy as

\be
\Gamma^{\mu}(q\approx 0) = \frac{\partial \Sigma (p)}{\partial p_{\mu}} 
\ee

Since the leading behavior near the Fermi surface is
\be
\Sigma (\omega,k) \propto (\omega-\mu) \ln|\omega-\mu|
\ee
\noindent  only the time component of the vertex would in principle lead to an infrared divergence, however, this vertex corresponds to the exchange of a longitudinal gluon which is Debye screened and does not lead to an infrared
divergence. The spatial components of the vertex are infrared finite, and do not change the leading infrared divergence.

The main conclusion extracted from this argument and based on the detailed analysis of reference\cite{ren} (to which we refer the reader)  is that
the infrared divergences in leading order in HDL {\em do not receive vertex corrections}. Hence we expect that the
resummation of the leading infrared divergences provided by the renormalization group either Euclidean or in real time
do not receive vertex corrections.

\section{Conclusions, and conjectures} 

Our goal in this article is a systematic study of the breakdown of the Fermi liquid description of the {\em normal} state
of cold and dense QED and QCD. As was recognized before, the exchange of magnetic gluons and photons that are only dynamically screened by Landau damping lead to infrared divergences in the quark self-energy for particles near the
Fermi surface. The main aspect of this article is the recognition that these divergences are akin to those arising in
a critical theory near its upper critical dimensionality. We then implemented both the Euclidean and the dynamical renormalization group to provide a non-perturbative resummation of the leading infrared singularities to analyze equilibrium and non-equilibrium aspects of the single (quasi) particle states near the Fermi surface. The non-perturbative resummation of the single particle propagator leads to a novel description of the spectrum of quasiparticle excitations with momentum
near the Fermi momentum summarized as follows

\begin{itemize}
\item[{\bf(i)}]{The quasiparticle pole, width and residue (wave function renormalization) are given by 
\bea
\tilde{\omega}_p(\tilde{k}) & = &  \mbox{sign}(\tilde{k}) \left[|\tilde{k}| M^{-2\lambda}\cos[\pi \lambda]  \right]^{\frac{1}{1-2\lambda}}\nonumber \\
 \Gamma(\tilde{k})  & = & Z_p[\tilde{k}] |\tilde{k}| \sin[\pi \lambda] \nonumber \\
 Z_p[\tilde{k}] & = & \frac{\left|\frac{\tilde{\omega}_p(\tilde{k})}{M} \right|^{2\lambda}}{(1-2\lambda)} \nonumber \\
\tilde{\omega}  & = & \omega-k_F ~~; ~~\tilde{k} = k-k_F \nonumber 
\eea

\noindent respectively. Therefore the residue of the ``quasiparticle pole'' and the ``quasiparticle width'' vanish near the fermi surface as

\bea
&&Z_p[\tilde{k}] \propto |k-k_F|^{\frac{2\lambda}{1-2\lambda}}  \\
&& \Gamma(\tilde{k})\propto |k-k_F|^{\frac{1}{1-2\lambda}}
\eea

\noindent and the group velocity of quasiparticles with Fermi momentum vanishes indicating a collective backflow. 

The real time evolution of the single quasiparticle wave function for states near the Fermi momentum is given by 

 \be
\psi_{k \approx k_F}(t) \approx \psi_{k \approx k_F}(t_0)  e^{-i[k_F+ \tilde{\omega}_p(\tilde{k})](t-t_0)}
e^{- \Gamma(\tilde{k})(t-t_0)}\left[\frac{t_0}{t} \right]^{2\lambda}  \nonumber
\ee
Therefore quasiparticles with the Fermi momentum relax with a power law determined by the  anomalous scaling dimension revealing that the
physics near the Fermi surface is similar to that of a critical theory. }

\item[{\bf (ii)}] {The single quasiparticle distribution function is {\em continuous} near the Fermi momentum, i.e. the
``jump'' discontinuity in the Fermi-Dirac distribution vanishes as a consequence of the vanishing of the quasiparticle
residue at the Fermi momentum. We find

\be
n_{k \approx k_F} = \frac{\sin[\pi \lambda]}{2\pi \lambda} - \frac{(k-k_F)}{\pi M(1-4 \lambda)} + {\cal O}(k-k_F)^2 
\ee

The vanishing of the discontinuity of the distribution function at the Fermi surface is the hallmark of the breakdown of
Fermi liquid theory. 
 }

\end{itemize}

There are some remarkable similarities between these features of cold dense QCD and QED and those of quasi one dimensional
Fermi systems with marginal interactions that lead to a description as Luttinger liquids\cite{schulz,shof}. These systems also
feature spectral densities and correlation functions with anomalous dimensions which are non-universal and depend on
 the couplings\cite{schulz,shof}. The fermion propagator that results from one gluon exchange (dynamically screened) is
also similar to that conjectured to describe the normal phase of high temperature superconductors in terms of marginal
Fermi liquids\cite{varma,anderson}. Even if these similarities are just coincidental, it is possible that a body of
results in the condensed matter literature could be useful to understand the fundamental aspects of the normal phase of
 QCD at high density.   

{\bf Potential impact of the results:} As we mentioned in the introduction, the properties of the normal state not only determine the
thermodynamics and transport properties but also influence those of the superconducting state, since superconductivity appears as an
instability of the normal state towards pairing interactions. Therefore the results obtained in this article that pertain solely to
the normal phase could bear on characteristics of the superconducting state as well as transport in the normal (and possible in the
superconducting) phase.
 
\begin{itemize}
\item[{\bf i):}] In reference\cite{ren} it was found that the same type of infrared divergences that lead to the breakdown of Fermi
liquid theory are responsible for a substantial decrease of the superconducting gap. In the calculation of reference\cite{ren}, these divergences are manifest through the
wave function renormalization in the equation for the gap. Furthermore, recently lifetime effects (of the normal state quasiparticles)
were shown to lead to a further decrease of the superconducting gap\cite{manuellifetime}. In this article we have shown how the
wave-function renormalization logarithmic divergence sums up to produce anomalous dimensions, which  change the quasiparticle
description. The spectral density features anomalous dimensions for frequency and momenta near the putative Fermi surface as well as
important modifications in the quasiparticle dispersion relations, lifetimes and residues. It is therefore a relevant question to assess the
potential corrections of the resummed spectral densities for the normal quark propagators to the superconducting gap. This entails solving the gap equation or alternatively the Dyson-Schwinger equation but with the normal quark propagators replaced by those obtained from the resummed spectral
functions. We are currently studying this possibility and expect to report on our findings in the near future. 

\item[{\bf ii):}] As mentioned in the introduction the most relevant physical setting for cold-dense QCD is that of astrophysical compact
objects, in particular protoneutron stars from Type II supernovae collapse or neutron stars (or pulsars). An important observational
aspect that could yield information from the core of neutron stars is cooling, which is studied through (soft)
 X-ray emmision\cite{schaab}. The cooling equation

$$
\frac{d E}{d t} = C \frac{dT}{dt}= -[L_{\nu}+L_{\gamma}] 
$$

\noindent relates the neutrino and photon luminosities to the specific heat $C$. 
 For normal quark matter the neutrino emissivity is dominated by the direct quark Urca processes $ d \rightarrow u~e~\overline{\nu}~~;~~
u~e \rightarrow d \nu$\cite{iwamoto}. 
A {\em normal} weakly interacting degenerate Fermi gas has a typical specific heat
linear in temperature $C(T) = C_0 T$ where the coefficient $C_0$ depends on the Fermi liquid properties. In particular the wave function
renormalization (residue at the quasiparticle pole) renormalizes the free field (free Fermi gas) value of $C_0$. Since the quasiparticle
residue vanishes, signaling the breakdown of Fermi liquid theory, we {\em conjecture} that the linear law will be replaced in a manner similar to that of Kondo-type systems\cite{mahan} $C \propto \ln(T)$. If a superconducting instability introduces a gap for all quarks,
then the specific heat will have a typical behavior $ C_1~ e^{-\Delta(T)/T}$ (for $T<T_c$) where $\Delta(T)$ is the superconducting gap and $C_1$ depends on the density of quasiparticle states near the Fermi surface which again receives non-Fermi liquid renormalization corrections. If the
neutrino and photon emissivity of quarks is suppressed because of the presence of a superconducting gap, then the electronic specific
heat will become relevant, in which case again the non-Fermi liquid corrections to the specific heat will be important since there are
no superconducting pairing instabilities in QED. If some of the quarks are un-gapped these are {\em normal} and
will receive renormalization of their Fermi liquid behavior.

\end{itemize} 

Understanding the potential corrections of the breakdown of Fermi liquid theory both for quarks and leptons could therefore lead to a
deeper understanding of cooling in neutron stars and therefore justifies further studies along these lines.  

 Our analysis was based on a perturbative approach, therefore its range of validity is restricted to asymptotically
large densities so that the effective coupling $\alpha_s(\mu)/6 \pi \ll 1$. Using the running of the QCD coupling constant,
the weak coupling condition implies chemical potentials and baryon densities orders of magnitude larger than those
available at the core of neutron stars. Thus our studies for the properties of the normal phase, along with those of color superconductivity must be taken as indicatives of novel and potentially relevant phenomena, but must be carefully extrapolated to the realm of baryochemical potential (baryon density) relevant to quark matter in neutron stars, requiring alternative non-perturbative techniques.

\acknowledgements

D.B. thanks NSF for support through grants PHY-9605186, PHY-9988720 and  CNRS-NSF-INT-9815064.
 H. J. d. V. thanks CNRS for support through a binational collaboration. The authors thank R. Pisarski and D. Rischke for fruitful discussions, D.B.
thanks D. Jasnow, K. Rajagopal, M. Stephanov,  D. Blaschke,  F. Weber, C. Roberts and S. Schmidt for illuminating discussions and comments.

%%%%%%%%%%%%%%%%%%%%%%%%%%%%%%%%

\appendix

\section*{Real-Time Propagators}
\subsection{Dirac Fields}

In this appendix we summarize the various
 real-time propagators used in this article.
The  fermion propagators  are
defined by
\begin{eqnarray}
&&\langle \Psi^{a}({\vec x},t) {\bar \Psi}^{b}({\vec x}',t')\rangle
= i \int \frac{d^3k}{(2\pi)^3} S_{\vec k}^{\,ab}(t,t')\,
e^{i{\vec k}\cdot({\vec x}-{\vec x}')},\nonumber\\
&&S_{\vec k}^{++}(t,t')=
S_{\vec k}^{>}(t,t')\theta(t-t')
+S_{\vec k}^{<}(t,t')\theta(t'-t), \nonumber\\
&&S_{\vec k}^{--}(t,t')=
S_{\vec k}^{>}(t,t')\theta(t'-t)
+S_{\vec k}^{<}(t,t')\theta(t-t'),\nonumber\\
&&S_{\vec k}^{\pm\mp}(t,t')=
S_{\vec k}^{\mbox{\scriptsize
\raisebox{1.8pt}{\raisebox{1.8pt}{$\scriptscriptstyle<$}
\raisebox{-1.5pt}{$\scriptscriptstyle\!\!\!\!\!\!\!\!\;>$}}}}(t,t'),
\label{fermionprop1}
\end{eqnarray}
where $a,\;b=\pm$.

In an equilibrium situation the propagators  can be written in terms of spectral densities as follows

\begin{eqnarray}
 iS_{\vec k}^{>}(t,t') & = & \int dq_0 \rho^>(q_0,\vec k) e^{-iq_0(t-t')} \nonumber \\
-iS_{\vec k}^{<}(t,t') & = & \int dq_0 \rho^<(q_0,\vec k) e^{-iq_0(t-t')} \nonumber \\
\rho^>(q_0,\vec k) & = & \rho(q_0,\vec k) (1-N_f(q_0,k)) \nonumber \\
\rho^<(q_0,\vec k) & = & \rho(q_0,\vec k) N_f(q_0,k) \label{specreps}
\end{eqnarray}
Where  the Fermi-Dirac distribution functions for particles and antiparticles, for the case under consideration of finite chemical potential $\mu$ and
zero temperature are given by 

\begin{eqnarray}
N_f(q_0) & = &  \Theta(\mu - q_0) \label{fermidistpart}  \\
\bar{N}_f(q_0) & = & \Theta(-\mu - q_0) \label{fermidistantipart}
\end{eqnarray}

For free fields  the fermion Wightman functions are given by 
\begin{eqnarray}
S_{\vec k}^{>}(t,t')&=&-\frac{i}{2\omega_{\vec k}}\Big\{
(\not\!{K}+m) [1-N_F(\omega_{\vec k})] e^{-i\omega_{\vec k}(t-t')}\nonumber\\
&&+\; \gamma_0(\not\!{K}-m)\gamma_0\, \bar{N}_F(\omega_{\vec k})\,
e^{i\omega_{\vec k}(t-t')}\Big\},\nonumber \\
S_{\vec k}^{<}(t,t')&=&\frac{i}{2\omega_{\vec k}}
\Big\{(\not\!{K}+m)\,N_F(\omega_{\vec k})\,e^{-i\omega_{\vec k}(t-t')}\nonumber\\
&&+\;\gamma_0(\not\!{K}-m)\gamma_0 [1-\bar{N}_F(\omega_{\vec k})]e^{i\omega_{\vec k}(t-t')}\Big\},
\label{fermionprop2}
\end{eqnarray}
with $K=(\omega_{\vec k},{\vec k})$, $\not\!{K}=\gamma^0\omega_{\vec k}-\vec{\gamma}\cdot \vec{k}$ and 
$\omega_{\vec k}= \sqrt{{\vec k}^2+m^2}$. For massless fermions (the case under consideration)

\begin{eqnarray}
\left\{\frac{(\not\!{K}+m)}{\omega_{\vec k}}\right\}_{m=0} & = & {\cal P}_-(\hat{\vec k})= \gamma^0-\vec{\gamma}\cdot  \hat{\vec k} \nonumber \\
\left\{\frac{\gamma_0(\not\!{K}-m)\gamma_0}{\omega_{\vec k}}\right\}_{m=0} & = & {\cal P}_+(\hat{\vec k})= \gamma^0+\vec{\gamma}\cdot \hat{\vec k} 
\label{proyectors}
\end{eqnarray}

with the properties

\begin{eqnarray}
({\cal P}_-(\hat{\vec k}))^2= ({\cal P}_+(\hat{\vec k}))^2 =0 \nonumber \\
{\cal P}_-(\hat{\vec k}){\cal P}_+(\hat{\vec k})= 2 \gamma^0{\cal P}_+(\hat{\vec k}) \nonumber \\
{\cal P}_+(\hat{\vec k}){\cal P}_-(\hat{\vec k})= 2 \gamma^0{\cal P}_-(\hat{\vec k}) \label{properties}
\end{eqnarray}

\subsection{Gauge Fields}
\subsubsection{Spatial components}

The transverse  photon propagators are defined by
\begin{eqnarray}
&&\langle A^{i,a}_T({\vec x},t) A^{j,b}_T({\vec x}',t')\rangle
=-i \int \frac{d^3q}{(2\pi)^3}{\cal G}_{T,q}^{ab}(t,t')\,
{\cal P}_T^{ij}({\vec q})\,e^{i{\vec q}\cdot({\vec x}-{\vec x}')},\nonumber\\
&&{\cal G}_{T,q}^{++}(t,t')=
{\cal G}_{T,q}^{>}(t,t')\theta(t-t')
+{\cal G}_{T,q}^{<}(t,t')\theta(t'-t),\nonumber \\
&&{\cal G}_{T,q}^{--}(t,t')=
{\cal G}_{T,q}^{>}(t,t')\theta(t'-t)
+{\cal G}_{T,q}^{<}(t,t')\theta(t-t'),\nonumber\\
&&{\cal G}_{T,q}^{\pm\mp}(t,t')=
{\cal G}_{T,q}^{\mbox{\scriptsize
\raisebox{1.8pt}{\raisebox{1.8pt}{$\scriptscriptstyle<$}
\raisebox{-1.5pt}{$\scriptscriptstyle\!\!\!\!\!\!\!\!\;>$}}}}(t,t'),
\label{gaugeprop1}
\end{eqnarray}
where
${\cal P}_T^{ij}({\vec q})=\delta^{ij}-\hat{q}^i \hat{q}^j$ is the
transverse projector and
the photon Wightman functions can be written in terms of a spectral representation as follows:

\begin{eqnarray}
{\cal G}_{T,q}^{>}(t,t')&=&i\int dq_0
\,\tilde{\rho}_T(q_0,q)\,[1+n_B(q_0)]\,e^{-iq_0(t-t')},\nonumber\\
{\cal G}_{T,q}^{<}(t,t')&=&i\int dq_0\,
\tilde{\rho}_T(q_0,q)\, n_B(q_0)\,e^{-iq_0(t-t')},
\label{wightmanT}
\end{eqnarray}
where $\tilde{\rho}_T(q_0,q)$ is the  spectral density and $n_B(q_0)$ is the Bose-Einstein distribution function. 

At zero temperature

\be
n_B(q_0)= -\Theta(-q_0) ~~; ~~ 1+n_B(q_0)= \Theta(q_0)
\ee

For free fields at zero temperature we find

\begin{eqnarray}
{\cal G}_{T,q}^{>}(t,t^{\prime})&=&
\frac{i}{2q}\,e^{-iq(t-t')},\nonumber\\
{\cal G}_{T,q}^{<}(t,t^{\prime})&=&
\frac{i}{2q}e^{iq(t-t')}\label{gaugeprop2}
\end{eqnarray}

In the HDL approximation, the spectral density $\tilde{\rho}_T(q_0,q)$ is given by

\begin{eqnarray}
\tilde{\rho}_{T}(q_0,q)& = & \mbox{sgn}(q_0)\,Z_{T}(q)\,\delta[q^2_0-\omega^2_{T}(q)]+\;\beta_{T}(q_0,q)\;
\theta(q^2-q_0^2)\nonumber \\
\beta_T(\omega,k)&=&\frac{\frac{g^2 \mu^2}{4\pi^2}\frac{\omega}{k}
\left(1-\frac{\omega^2}{k^2}\right)}
{\left\{\omega^2-k^2-\frac{g^2 \mu^2}{4\pi^2}
\left[\frac{2\omega^2}{k^2}+\frac{\omega}{k}
\left(1-\frac{\omega^2}{k^2}\right)
\ln\left|\frac{k+\omega}{k-\omega}\right|\right]\right\}^2+
\left[\frac{ g^2 \mu^2}{4\pi}\frac{\omega}{k}
\left(1-\frac{\omega^2}{k^2}\right)\right]^2}. 
\label{tilderhot}
\end{eqnarray}
and $Z_{T}(q)$ is the residue at the pole of the collective excitation\cite{lebellac}.

\subsubsection{Time component of the  photon propagator}

The time-time component of the photon propagator describes the instantaneous Coulomb interaction
including screening corrections and are given by 
\begin{eqnarray}
&&\langle A^{a}_0({\vec x},t) A^{b}_0({\vec x}',t')\rangle
= i \int \frac{d^3q}{(2\pi)^3}\,{\cal G}_{L,q}^{ab}(t,t')
e^{i{\vec q}\cdot({\vec x}-{\vec x}')},\nonumber\\
&&{\cal G}_{L,q}^{++}(t,t')= \frac{1}{q^2}\delta(t-t')+
{\cal G}_{L,q}^{>}(t,t')\theta(t-t')
+{\cal G}_{L,q}^{<}(t,t')\theta(t'-t),\nonumber \\
&&{\cal G}_{L,q}^{--}(t,t')= -\frac{1}{q^2}\delta(t-t')+
{\cal G}_{L,q}^{>}(t,t')\theta(t'-t)
+{\cal G}_{L,q}^{<}(t,t')\theta(t-t'),\nonumber \\
&&{\cal G}_{L,q}^{\pm\mp}(t,t')=
{\cal G}_{L,q}^{\mbox{\scriptsize
\raisebox{1.8pt}{\raisebox{1.8pt}{$\scriptscriptstyle<$}
\raisebox{-1.5pt}{$\scriptscriptstyle\!\!\!\!\!\!\!\!\;>$}}}}(t,t'),
\end{eqnarray}
with the Wightman functions expressed in terms of the
spectral density $\tilde{\rho}_L$ as
\begin{eqnarray}
{\cal G}_{L,q}^{>}(t,t')&=& -i\int dq_0
\,\tilde{\rho}_L(q_0,q)\,[1+n_B(q_0)]\,e^{-iq_0(t-t^\prime)},\nonumber\\
{\cal G}_{L,q}^{<}(t,t')&=& -i\int dq_0\,
\tilde{\rho}_L(q_0,q)\,n_B(q_0)\,e^{-iq_0(t-t^\prime)}.
\label{wightmanL}
\end{eqnarray}

For free fields $\tilde{\rho}_L(q_0,q)=0$ and 

\begin{eqnarray}
&&{\cal G}_{L,q}^{++}(t,t')= \frac{1}{q^2}\delta(t-t'),\nonumber \\
&&{\cal G}_{L,q}^{--}(t,t')= -\frac{1}{q^2}\delta(t-t'),\nonumber \\
&&{\cal G}_{L,q}^{\pm\mp}(t,t')= 0 \label{freelongi}
\end{eqnarray}

In the Hard Dense Loop (HDL) approximation the spectral density $\tilde{\rho}_{L}(q_0,q)$ is given by

\begin{eqnarray}
\tilde{\rho}_{L}(q_0,q)&=&\mbox{sgn}(q_0)\,Z_{L}(q)\,
\delta[q^2_0-\omega^2_{L}(q)]
+\;\beta_{L}(q_0,q)\;\theta(q^2-q_0^2)\;, \nonumber\\
\beta_{L}(q_0,q)&=&
\frac{\frac{g^2\mu^2}{2\pi^2}\frac{q_0}{q}}{\left[q^2+
\frac{g^2\mu^2}{2\pi^2}\left(2-\frac{q_0}{q}\ln\left|\frac{q+q_0}{q-q_0}\right|\right)\right]^2+
\big[\frac{g^2 \mu^2}{2\pi}\frac{q_0}{q}\big]^2}\;,\label{rholong}
\end{eqnarray}
where $\omega_{L}(q)$ is the plasmon (longitudinal photon) pole
and $Z_{L}(q)$ is the corresponding residue~\cite{lebellac}.

%%the bibliography goes here

%\input{nflqcdbiblio}

%%%%%%%%%%bibliography begins here 

%%%%%%%%%%% BEGIN FIGURES %%%%%%%%
%%%%%%%%%%%%begin figures%%%%%%%%%%%%%

%%%%%%%%begin figure 1 %%%%%%%%%%%%%
%%%%modulus squared of mode function at t_NL for broken symmetry

\begin{figure}
\centerline{ \epsfig{file=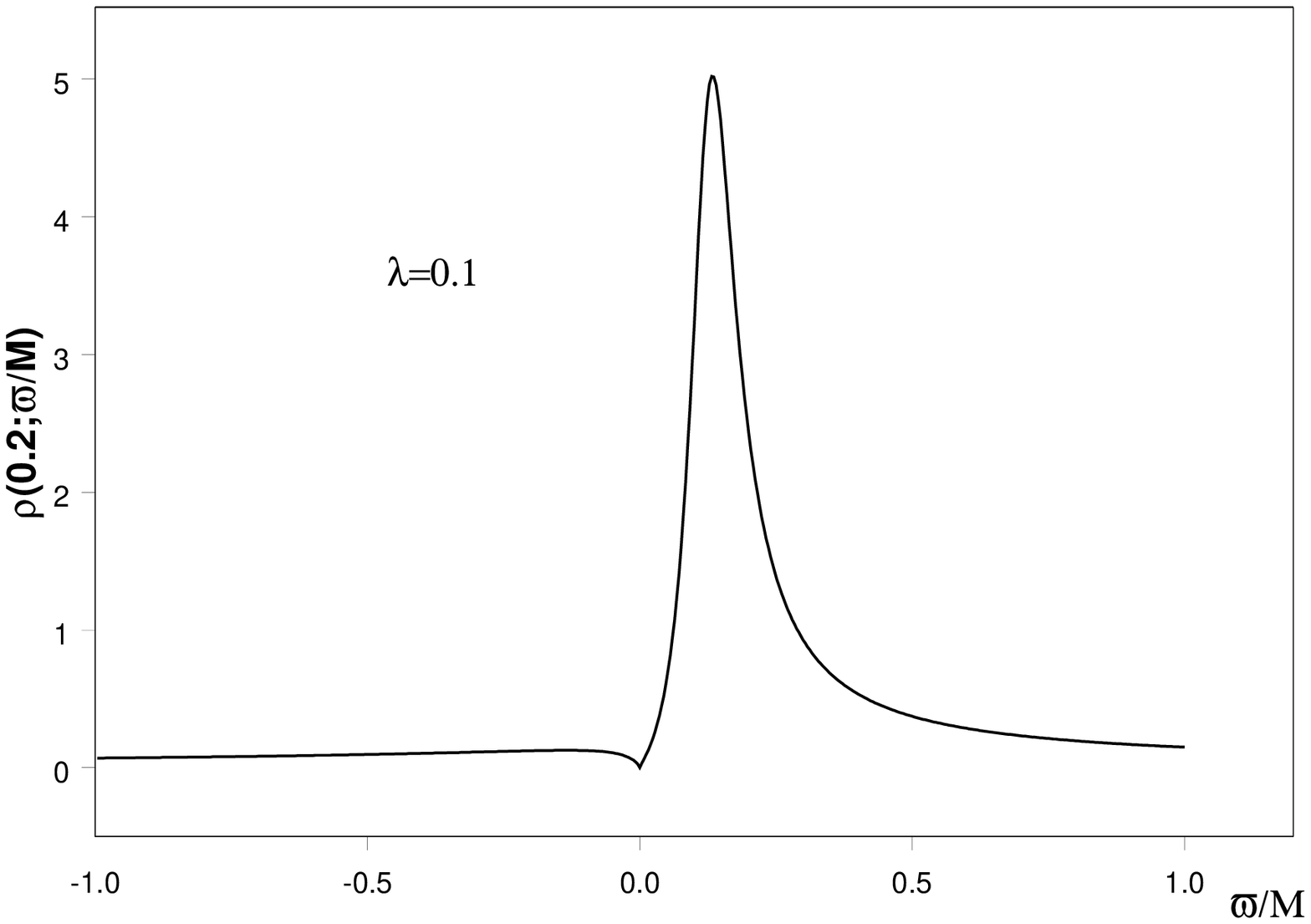,width=7in,height=7in}}
\caption{$\rho_-(\frac{\tilde{k}}{M},\frac{\tilde{\omega}}{M})$ for $ \lambda=0.1;\frac{\tilde{k}}{M}=0.2 $.  \label{fig1}} 
\end{figure}

%%%%%%%%end figure 1%%%%%%%%%%%

\newpage

%%%%%%%%begin figure 2 %%%%%%%%%%%%%
%%%%modulus squared of mode function at t_NL for broken symmetry

\begin{figure}
\centerline{ \epsfig{file=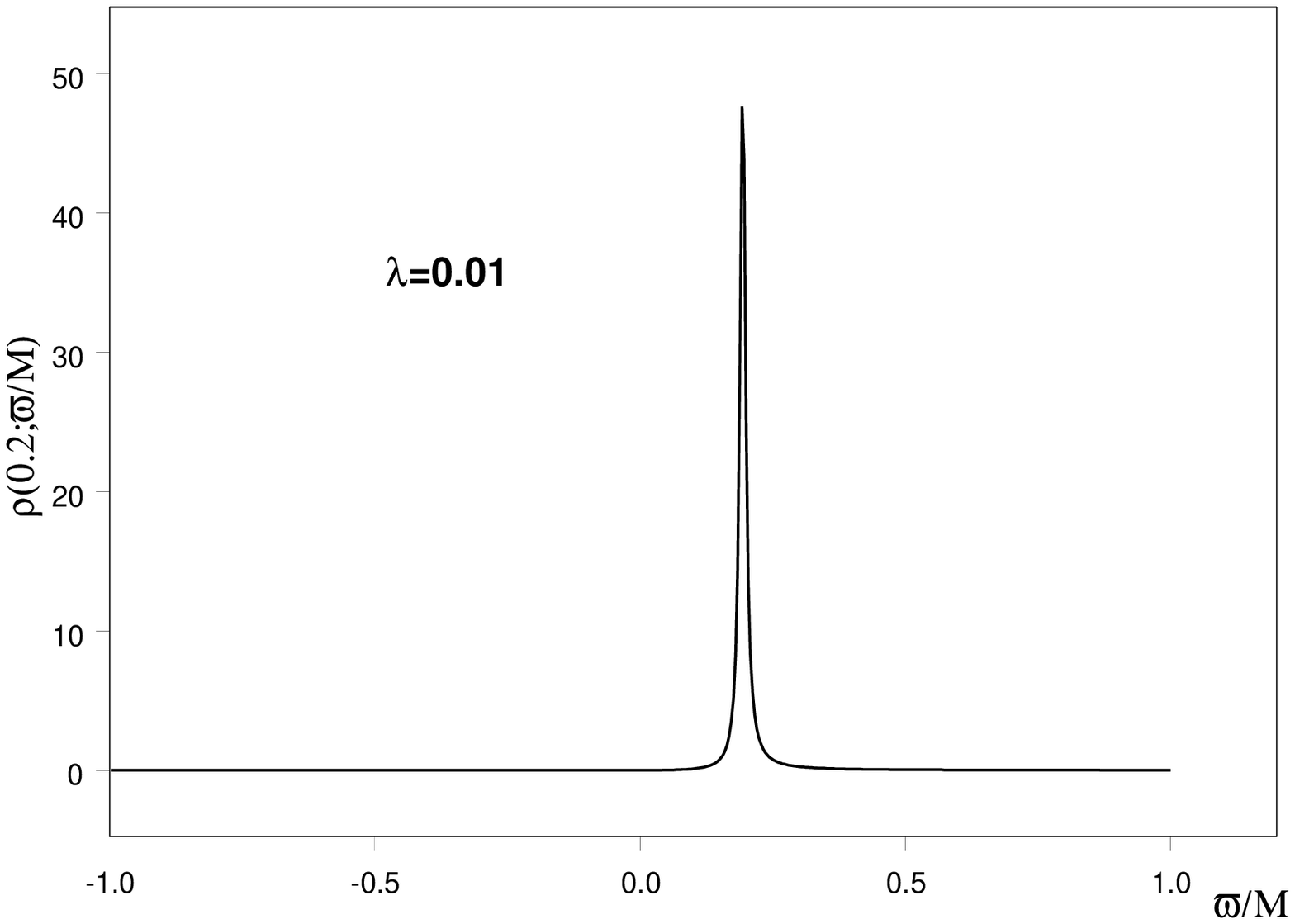,width=7in,height=7in}}
\caption{$\rho_-(\frac{\tilde{k}}{M},\frac{\tilde{\omega}}{M})$ for $ \lambda=0.01;\frac{\tilde{k}}{M}=0.2 $.  \label{fig2}} 
\end{figure}

%%%%%%%%end figure 2%%%%%%%%%%%

%%%%%%%%% END FIGURES %%%%%%%%%%%%%%

\end{document}